\documentclass[11pt,a4paper]{article}
\pdfoutput=1
\usepackage{jheppub,bm,booktabs,multirow}
\usepackage{xcolor}
\usepackage{overpic}
\usepackage{verbatim}

\allowdisplaybreaks

\makeatletter
\def\@fpheader{~}
\makeatother

\usepackage[Q=yes,pverb-linebreak=no]{examplep}

\def\nno{\nonumber}

\title{Resummation of non-global logarithms in cross sections with massive particles}
\author[a]{Marcel Balsiger}
\author[a]{\!, Thomas Becher}
\author[b]{and Andrea Ferroglia}

\affiliation[a]{Albert Einstein Center for Fundamental Physics, Institut f\"ur Theoretische Physik, Universit\"at Bern,
  Sidlerstrasse 5, CH-3012 Bern, Switzerland}
  \affiliation[b]{Physics Department, New York City College of Technology, The City University of New York, 300 Jay Street, Brooklyn, NY 11201 USA}

\emailAdd{balsiger@itp.unibe.ch}
\emailAdd{becher@itp.unibe.ch}
\emailAdd{aferroglia@citytech.cuny.edu}

\date{\today}

\abstract{A factorization formalism for jet processes involving massive colored particles such as the top quark is developed, extending earlier results for the massless case. The factorization of soft emissions from the underlying hard process is implemented in an effective field theory framework, which forms the basis for the resummation of large logarithms. The renormalization group evolution giving rise to non-global logarithms is implemented into a parton shower code in the large-$N_c$ limit. After a comparison of the massive and massless radiations patterns, the cross section for $t\bar{t}$ production with a veto on additional central jet activity is computed, taking into account radiation both from the production and the decay of the top quarks.  The resummation of the leading logarithms leads to an improved description of ATLAS measurements at $\sqrt{s}=7\,{\rm TeV}$.}

\begin{document}

\maketitle

\section{Introduction}

The study of jet cross sections plays a crucial role in high-energy physics. While theoretical calculations are carried out in terms of interactions at the field level, detectors are only able to measure properties of outgoing particles after they have fully hadronized, i.e. transformed from colored quarks and gluons to color-neutral final states such as mesons. Consequently, it is impossible to measure the underlying hard scattering process directly, but one needs to reconstruct it by measuring jets and analyzing their properties.

While the total energy of the particles inside the jets is typically of the same order as the partonic center-of- mass energy of the collision, the total energy of the particles not ending up in a jet is considerably lower. 
Due to this scale separation effective field theory methods, in particular Soft-Collinear Effective Theory (SCET) \cite{Bauer:2000yr,Bauer:2001yt,Beneke:2002ph} (see \cite{Becher:2014oda,Becher:2018gno,Cohen:2019wxr} for reviews), are useful in the study of jet cross sections.
In the effective theory the cross sections factor into hard, collinear and soft functions, each of which can be safely evaluated in fixed-order perturbation theory at their characteristic energy scale. To connect these factors, it is then necessary to evolve one of the factors from its characteristic scale to the scale of the other factor by using the Renormalization Group (RG) equation. This procedure was first applied to jet cross sections in \cite{Becher:2015hka,Becher:2016mmh}.  

Because of their multi-scale nature, jet cross sections are sensitive to potentially large logarithmic corrections. When evaluating the phase-space integrals of matrix elements and applying cuts to the allowed energies,  logarithms of the ratios of the energy scales involved in the process appear in the calculations. For example, when the energy of particles inside the jets (denoted by $Q$) is unconstrained and of the same order as the partonic center-of-mass energy, i.e.~$Q^2\sim\hat{s}$, but the energy outside the jets (denoted by $Q_0$) is required to be small, the phase-space integrals produce terms proportional to $\ln\left(Q/Q_0\right)$. These logarithms become large if $Q_0 \ll Q$.

The factorization formula studied in \cite{Becher:2015hka,Becher:2016mmh}, derived within the effective field theory approach, can be used to resum these corrections, in principle to all logarithmic orders. Based on this theoretical framework, a dedicated parton shower code was developed and applied to resum the large logarithms appearing in jet processes and isolation-cone cross sections up to leading logarithmic (LL) order  in \cite{Balsiger:2018ezi}. Subsequently, higher-order matching corrections in both the hard and the soft function were added. This led to the resummation of the interjet energy flow up to LL$^\prime$ accuracy and to the resummation of the jet mass up to next-to-leading-logarithmic (NLL$^\prime$) accuracy in \cite{Balsiger:2019tne}. As usual, the prime in LL$^\prime$ and NLL$^\prime$ indicates that the matching corrections are included one order higher than what it would be required in RG improved perturbation theory. In the present case, this means that NLO hard and soft functions were used. By supplementing these calculations with the two-loop corrections to the anomalous dimension matrix one would achieve full NLL and next-to-next-to-leading logarithmic (NNLL) accuracy, respectively.  

The work done so far was carried out in the high-energy limit where all partons can be considered massless. The purpose of this paper is to extend the approach of \cite{Becher:2015hka,Becher:2016mmh,Balsiger:2018ezi,Balsiger:2019tne} to processes involving heavy colored particles and to develop and validate a  parton shower code  for the resummation of jet cross sections in top-quark production processes. Soft radiation is obtained from matrix elements of Wilson line operators along the directions of the emitting particles, independently of the mass of the emitting parton. Because of this fact, the factorization theorem has the same general form as in the purely massless case. However, the soft radiation pattern and its generation by the parton shower code differ significantly in the two cases. At one-loop order, the angular dependence of the radiation of a soft parton with momentum $k^\mu = E\, n_k^\mu$ between legs carrying momenta $p_i$ and $p_j$ is given by the usual product of eikonal factors
\begin{align}
W_{ij}^k=\frac{p_i \cdot p_j}{p_i\cdot n_k \,\, n_k\cdot p_j} \label{eq:radfact} \, . 
\end{align}
This factor is the same in both cases, but massless particles are traveling along light-like directions, while massive particles travel along time-like directions. This difference in kinematics must be accounted for in the shower code.  Furthermore,  in contrast to what happens in the high-energy limit, the radiation factor in (\ref{eq:radfact}) does not vanish when $i=j$,  if $p_i$ is a time-like momentum. Therefore, in addition to the usual dipole emission pattern, it is necessary to include monopole contributions in the massive case. The latter describes radiation  that is emitted and absorbed by the same Wilson line rather than exchanged between two color-connected Wilson lines. This difference in the massive and massless radiation pattern is of course well known, in particular the different collinear behavior, which is often referred to as the {\em dead cone effect} \cite{Dokshitzer:1991fc,Dokshitzer:1991fd,Ellis:1991qj,Maltoni:2016ays}.

\begin{figure}[t!]
	\centering
	\begin{overpic}[scale=0.30]{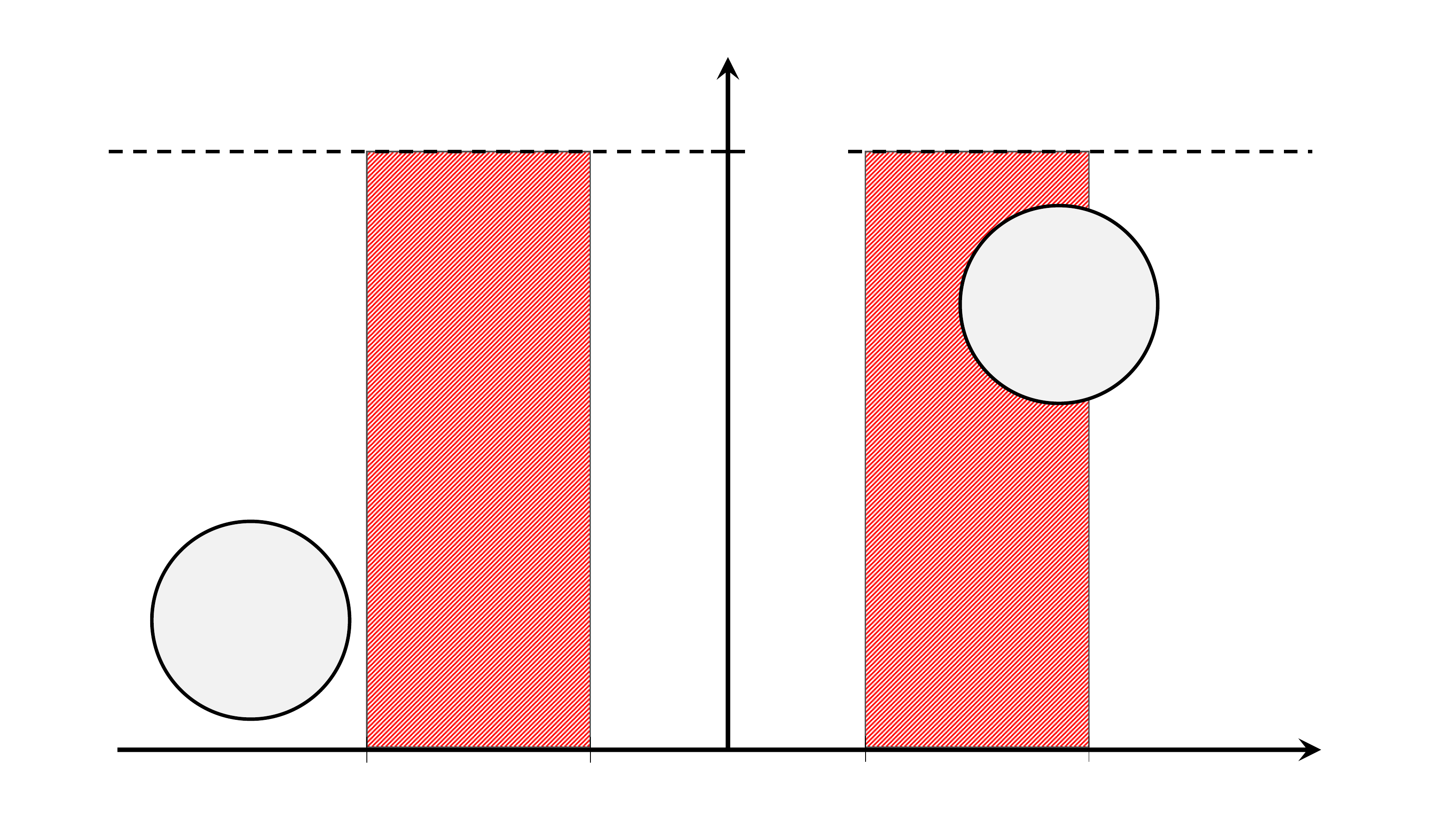}
		\put(49.2,54.5){$ \phi$}
		\put(92.5,4.2){$ y$}
		\put(58,1){$ y_{\rm min} $}
		\put(36.5,1.0){$ -y_{\rm min} $}
		\put(73.8,1){$y_{\rm max} $}
		\put(21,1){$-y_{\rm max} $}
		\put(14,12.5){$b$-jet}
		\put(70,34){$\bar{b}$-jet}
		\put(52.6,45.0){$2\pi$}
	\end{overpic}
	\caption{Sketch of the veto region as defined by ATLAS in \cite{ATLAS:2012al}. The gap, in which additional radiation is vetoed, is represented by the shaded red area with rapidity $y_{\rm min} <|y|<y_{\rm max} $. Radiation inside the $b$-tagged jets is not vetoed. For $y_{\rm min}=0$, this setup reduces to the usual central jet veto.\label{fig:outsideRegion}}
\end{figure}
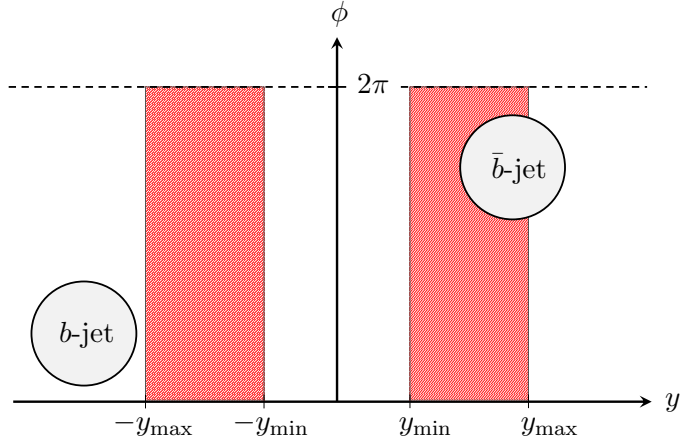

As an application of the new parton shower code described in this work, we consider $t\bar{t}$ production with a veto on additional central jet energy. This process was measured by ATLAS at the Large Hadron Collider (LHC) with the goal of testing the description of soft radiation in parton showers \cite{ATLAS:2012al}. The top pair production process involves two initial-state partons producing a $t \bar{t}$-pair in the final state. The top quarks then decay into bottom quarks and $W$ bosons. The measurement is performed using events in which the $W$'s decay leptonically and in which two $b$-jets are detected. The veto on central jets is imposed by requiring that, with the exception of the two bottom-tagged jets, no additional jets above a given transverse momentum $Q_0$ are allowed to be present in the rapidity range $y_{\rm min} <|y|<y_{\rm max} $  (see Figure \ref{fig:outsideRegion}).  With the veto, only particles of low energy  are allowed inside this rapidity range, while the energy is unconstrained anywhere else. This is a typical situation in which large non-global logarithms appear. In this work these logarithms are resummed  at LL accuracy and the results of the resummation are matched to NLO predictions in fixed-order perturbation theory. 

\begin{figure}[t!]
	\centering
	\begin{overpic}[width=0.75\textwidth]{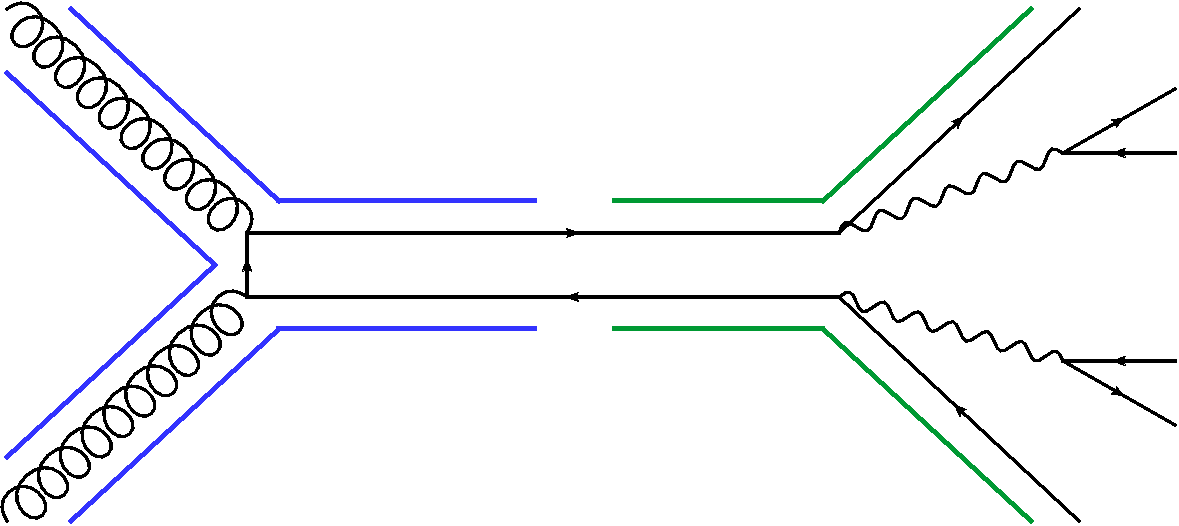}
		\put(93,45){$ b$}
		\put(93,-2){$ \bar{b}$}
		\put(86,26){$ W^+$}
		\put(86,17){$ W^-$}
		\put(101,37){$ \nu$}
		\put(101,30){$ l^+$}
		\put(101,13){$ \bar{\nu}$}
		\put(101,7){$ l^-$}
		\put(48,26.5){$ t$}
		\put(48,15){$ \bar{t}$}
		\put(-2,45){$ g$}
		\put(-2,-2){$ g$}
	\end{overpic}
	\caption{Diagram for the process $gg\rightarrow t\bar{t}\rightarrow b\bar{b}\,l ^+ \, l^- \, \nu\bar{\nu}$. In the large-$N_c$ limit, the radiation can be split into a set of color dipoles. The color dipoles associated to the production of the $t\bar{t}$ pair are shown in blue, the ones associated to the decay in green. The full LL cross section will include the emissions from all five dipoles.\label{fig:dipolestructure}}
\end{figure}
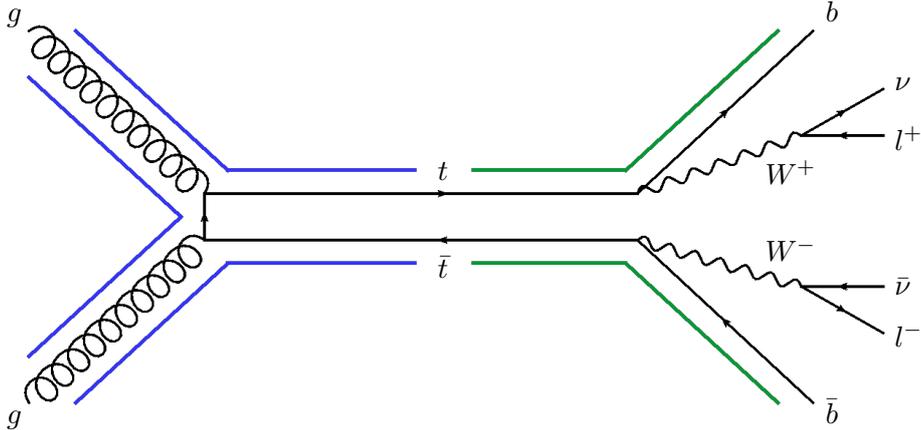

In addition to radiation effects associated with the production process, one should also include radiation emerging from the decay products of the top quarks. We work in the narrow-width approximation for the top quarks, in which they are treated as stable particles and the process factorizes into a production cross section multiplied by the decay of the top quarks. It is well known that radiation from the $b$-quarks that would contribute to non-factorizable corrections in fixed-order perturbation theory is suppressed by factors of $\mathcal{O} (\Gamma_t/m_t)$ \cite{Fadin:1993dz,Fadin:1993kt,Melnikov:1993np,Melnikov:1995fx,Beenakker:1999ya,Denner:1997ia}. To account for the factorizable contributions, we run a separate shower for the top decay to also account for the $b$-quark radiation. Numerically, the effect of this radiation is
smaller than the one from the production of the top pair since the radiation inside the $b$-jet is not constrained. However, the radiation from the decay is large enough that it must be taken into account. Figure \ref{fig:dipolestructure} shows one of the several tree-level diagrams contributing to the  $t\bar{t}$-pair production process  measured by ATLAS in \cite{ATLAS:2012al}. We also depict the color dipoles, which are the source of the emissions in the large-$N_c$ limit.  

The remainder of this paper is organized as follows. In Section \ref{sec:factorization}  the factorization theorem \cite{Becher:2015hka,Becher:2016mmh}  is reviewed and the changes needed in presence of  massive partons  are discussed. Section \ref{sec:evaluation} describes  in detail how the relevant phase-space integrals can be evaluated in the parton shower code. In Section \ref{sec:massiveeffects} we assess  the impact of massive partons in the resummation of non-global logarithms. An explicit example of the resummation of non-global logarithms for a cross section involving top quarks at LL accuracy is presented in Section \ref{sec:resummation}. As indicated above, the observable  we consider is top-pair production with a veto on central jet energy. The predictions for this observable are then matched to the NLO result and compared  to experimental measurements carried out by ATLAS. Section \ref{sec:conclusion} contains our conclusions and an outlook. 
 In Appendix~\ref{sec:MCalg}, we use a sample event to illustrate our parton shower code step by step. In Appendix \ref{sec:FOofLL} we explain how to use the shower to also compute the first two orders of the fixed-order expansion of the resummed result.

\section{Factorization for cross sections involving massive quarks} \label{sec:factorization}

Before discussing the factorization of the cross section, we should determine  which scales are present  and which scale hierarchies can arise in the observable under study. Throughout this paper, we consider scattering processes at a large center-of-mass energy $Q$ and impose a veto on radiation in a certain phase-space region. We are interested in a regime where the energy scale $Q_0$ of the soft radiation in the veto region is much smaller than $Q$. The presence of the massive particles introduces additional scales in the process. On top of the masses themselves, which we denote generically with $M$, the most important new scale is the production threshold: when massive particles are produced, only part of the energy $Q$ is available for  additional radiation.   We  denote by $Q_1$ the energy above threshold that can be radiated. For $t\bar{t}$ production, the threshold is at $Q_T=2m_t$ and the maximal energy available to radiation is 
\begin{equation}\label{eq:hardest}
Q_1 = \frac{Q^2-4m_t^2}{2Q}\, .
\end{equation}
Kinematically, this value corresponds to a configuration where a collinear $t\bar{t}$ pair recoils against a gluon. This is a corner of phase space and the typical gluon energy will be much lower. However, the scale of the hardest possible emission plays an important role since it corresponds to the large scale in the emission process which should be compared to the veto scale $Q_0$. Since we are interested in non-global logarithms associated with soft radiation, we only consider $Q_1 \gg Q_0$, but even under this assumption, one should consider two different hierarchies, namely a) $Q \sim Q_1$ and b) $Q \gg Q_1$.

The simpler of the two cases is $Q \sim Q_1$, which implies that the process energy is far larger than the  threshold energy, and that the masses are smaller than the maximum emission energy, $Q_1\gg M$. It is then interesting to ask what role the masses themselves play and whether we encounter logarithms of the masses. If the heavy partons are not in the veto region, the vetoed cross section is collinear safe and mass effects are power suppressed in the limit $M \to 0$; the massless limit is smooth. On the other hand, if the massive partons are inside the veto region, the limit $M \to 0$  becomes complicated. Of course, in top-pair production, several additional complications arise when considering the limit $m_t \to 0$. In this paper, we only consider the case where $M$ is larger or of the same order as $Q_0$. 

In the case in which $Q \gg Q_1$ instead, the process occurs near threshold and the emissions are always soft compared to the particle masses. At the same time, we want to have $Q_1 \gg Q_0$, therefore the distance of $Q$ from the threshold must still be large compared to the scale of soft radiation. Phenomenologically, this situation can only be relevant for top quarks and quite stringent vetoes. Since the radiation is always soft compared to the heavy particles, we should describe the entire process in Heavy Quark Effective Theory (HQET) (for reviews see \cite{Neubert:1993mb,Manohar:2000dt}). One would first match QCD onto HQET at the scale $Q$ and evolve to $Q_1$ before computing the soft emissions. It should be noted that this first matching will also have to be performed for the total cross section in the same kinematic regime. The effect will therefore largely cancel in ratios of cross sections such as the gap fraction. Furthermore,  if one gets very close to the threshold $Q_1 \sim M \alpha_s$, the heavy quarks become non-relativistic, but in view of $Q_1 \gg Q_0$ this regime is not important phenomenologically. When we apply our formalism to top-pair production at the LHC at $\sqrt{s}= 7\,{\rm TeV}$, we find that the average $Q$ of the partonic collisions is  $Q\approx 520 \,{\rm GeV} \sim 3 m_t$ and $Q_1 \approx 150 \,{\rm GeV}$. Therefore, in the phenomenological application considered in this work the scale hierarchy lies in between cases  a) and b).

We discussed the two scenarios a) and b), but together with the scale $M$, also combinations of scenarios can arise. For example, for $Q_1 \gg M$, it is possible to emit additional massive partons (at leading logarithmic accuracy only gluons are emitted). Then, for $Q_1 \gg M \gg Q_0$, one could imagine a two step procedure, where one would start in scenario a), but after a number of emissions, only a small energy is left and one would switch over to scenario b). Massive theories have a much richer set of kinematic configurations than massless ones and can involve complicated interplays of different scales. 

Here we first describe factorization for the simple case a), restricting ourselves to $e^+e^-$ cross sections for the moment. After presenting the result, we discuss how it should be modified to account for the case b). The factorization formula for a jet production with a veto on radiation in part of the phase space takes the form
\begin{align}\label{eq:Q_1}
\sigma(Q, Q_0) &=  \sum_{m=m_0}^\infty \big\langle \bm{\mathcal{H}}_m(\{\underline{v}\},\{\underline{n}\},Q,\mu) \otimes \bm{\mathcal{S}}_m(\{\underline{v}\},\{\underline{n}\},Q_0,\mu) \big\rangle\, ,
\end{align}
where $m_0$ is the number of final-state jets. The hard function $\bm{\mathcal{H}}_m$ describes the production of $m$ partons in the unconstrained region and the soft function $\bm{\mathcal{S}}_m$ is the matrix element squared of the emission from Wilson lines along the $m$ partons of the hard function. Both of these functions depend on the directions of the $k$ massive partons $\{\underline{v}\}=\{v_{1},\dots,v_k\}$ and $m-k$ massless partons $\{\underline{n}\}=\{n_{k+1},\dots,n_m\}$. As discussed above, the hard functions $\bm{\mathcal{H}}_m$ also depend on the particle masses and derived quantities such as $Q_1$. In order not to overburden the notation, we suppress this dependence. The symbol $\otimes$ indicates an integral over the directions of the $m$ particles and $\langle \,\dots\, \rangle$ denotes the color trace which is taken after combining the functions. Up to the fact that some reference vectors are time-like, this formula is identical to the one studied in \cite{Becher:2016mmh,Balsiger:2018ezi,Balsiger:2019tne}.

The hard functions $\bm{\mathcal{H}}_m$ are free of large logarithms if one chooses a value $\mu \sim Q$ for the renormalization scale. The same is achieved for the soft functions $\bm{\mathcal{S}}_m$ for $\mu \sim Q_0$. For $Q \gg Q_0$, at least one of these two functions will involve large logarithms, irrespective of the scale choice. These large logarithms can be resummed by solving the RG equation of the hard function and evolving it from its characteristic scale $\mu_h \sim Q$ down to a soft scale  $\mu_s \sim Q_0$, leading to 
\begin{align} \label{eq:crssctEvo}
\sigma(Q, Q_0) &= \sum_{l=m_0}^\infty \big\langle \bm{\mathcal{H}}_l(\{\underline{v}\},\{\underline{n}^\prime\},Q,\mu_h) 
\otimes \sum_{m\geq l}^\infty \bm{U}_{lm}(\{\underline{v}\},\{\underline{n}\},\mu_s,\mu_h)\,\hat{\otimes}\, 
\bm{\mathcal{S}}_m(\{\underline{v}\},\{\underline{n}\},Q_0,\mu_s) \big\rangle \,,
\end{align}
where the evolution factor is just the path-ordered exponential of the anomalous dimension 
\begin{align}
\bm{U}(\{\underline{n}\},\mu_s,\mu_h) = {\rm \bf P} \exp\left[ \int_{\mu_s}^{\mu_h} \frac{d\mu}{\mu} \bm{\Gamma}^{H}(\{\underline{v}\} , \{\underline{n}\},\mu) \right], \label{eq:evolmat}
\end{align}
which evolves the $l$-parton configuration along the time-like directions $\{\underline{v}\}=\{v_{1},\dots , v_k\}$ and the light-like directions $\{\underline{n}^\prime\}=\{n_{k+1},\dots,n_l\}$ into an $m$-parton final state  including  the time-like directions $\{\underline{v}\}$ and the light-like directions $\{\underline{n}\}=\{n_{k+1},\dots,n_l,n_{l+1},\dots,n_m\}$. RG evolution generates additional massless particles and $\hat{\otimes}$ denotes the integration over their directions before integrating over the 
hard directions. 

Up to now we worked under the assumptions that $Q_1 \sim Q$. Alternatively, if $Q\gg Q_1$, the hard functions involve large logarithms of the ratio $Q_1/Q$ which are not resummed by the above treatment. In order to factorize the two scales, one must first match onto HQET. For $e^+ e^- \to \gamma^* \to t \bar{t}$, it is necessary  to match the electromagnetic current operator which induces the process onto the corresponding HQET operator
\begin{equation}\label{eq:HQET}
J^\mu_{\rm e.m.} = \bar{t} \,\gamma^\mu \, t  \to  C_V(v\cdot v' ,Q, \mu) \bar{h}_{v'} \gamma^\mu h_v \,,
\end{equation}
where $h_v$ and $h_{v'}$ are the two HQET fields describing the top and the anti-top quarks. One would then derive an expression analogous to \eqref{eq:crssctEvo} in HQET. The hard functions arising would be related to the ones in \eqref{eq:crssctEvo} by
\begin{equation}
\bm{\mathcal{H}}_l(\{\underline{v}\},\{\underline{n}^\prime\},Q,\mu) = \left|C_V(v\cdot v' ,Q, \mu)\right|^2 \, \bm{\mathcal{H}}^{\rm HQET}_l(\{\underline{v}\},\{\underline{n}^\prime\},Q_1,\mu)  + \mathcal{O}(Q_1/Q)\,.
\end{equation}
To resum the logarithms of $Q_1/Q$, one will first solve the RG of $C_V$ to run from the scale $\mu \approx Q$ down to $\mu \approx Q_1$. When computing the gap fraction, one will also compute the total cross section in HQET using \eqref{eq:HQET}. The anomalous dimension driving the running of $C_V$ is the massive cusp anomalous dimension with the cusp angle defined by the two vectors $v$ and $v'$. In the ratio defining the gap fraction, the Wilson coefficient $C_V$ and its running will drop out. The situation is more complicated for hadron colliders, which involve sums of different partonic channels with different running so that the cancellation between numerator and denominator will not be complete. The general form of the anomalous dimension for a process with massive partons was given in \cite{Becher:2009kw} and the explicit forms relevant for top production can be found in \cite{Ahrens:2010zv}, but we will not study the small effect of this running in this work. However, an important lesson from the above discussion is that one should set the scale $\mu_h\sim Q_1$ in observables such as the gap fraction, since most of the running above this scale will drop out in the ratio of cross sections.

In \eqref{eq:evolmat} we have presented the formal solution to the evolution equation. We will now discuss how the general solution simplifies at LL accuracy and how it can be implemented as a parton shower. 
In dijet processes at lepton colliders, one only needs to consider the case $l=m_0=2$ at LL,  as the contribution of additional partons to the hard function would be suppressed by additional powers of $\alpha_s$ for $\mu_h\sim Q$. On the other side of the energy spectrum, the LL soft function is simply the unit matrix in the color space of the $m$ final-state partons, since any soft correction would again be suppressed by a factor $\alpha_s$ at the low scale $\mu_s \sim Q_0$. When computing $t\bar{t}$ production at a future electron-positron collider with a sufficiently high center-of-mass energy at LL accuracy, the general result  \eqref{eq:crssctEvo} therefore simplifies to
\begin{align} 
\sigma_{\text{LL}}(Q, Q_0) &=\sum_{m=2}^\infty \big\langle \bm{\mathcal{H}}_2(\{v_1,v_2\},\{\},Q,\mu_h) 
\otimes  \bm{U}_{2m}(\{v_1,v_2\},\{\underline{n}\},\mu_s,\mu_h)\,\hat{\otimes}\, 
\bm{1} \big\rangle\,.
\end{align}
The situation is more complicated at hadron colliders such as the LHC, where the initial state contains two additional colored hard partons, which give rise to non-perturbative Parton-Distribution Functions (PDFs). In addition, Glauber gluons can induce interactions between soft and collinear partons. This complication is absent in the large-$N_c$ limit in which we perform our computations. In this limit, the only difference to the $e^+e^-$ case is that there are two additional Wilson lines which describe the soft initial-state radiation. 

For LL resummation one needs the anomalous dimension only at one-loop accuracy. Consequently, the exponent of the evolution matrix in (\ref{eq:evolmat}) reduces to
\begin{equation}\label{eq:evolutiontime}
\int_{\mu_s}^{\mu_h} \frac{d\mu}{\mu}\, \bm{\Gamma}^H= \int_{\alpha_s(\mu_s)}^{\alpha_s(\mu_h)} \frac{d\alpha}{\beta(\alpha)}\, \frac{\alpha}{4\pi}\,\bm{\Gamma}^{(1)}=  \frac{1}{2\beta_0}\ln\frac{\alpha_s(\mu_s)}{\alpha_s(\mu_h)} \,\bm{\Gamma}^{(1)}  \equiv t \,\bm{\Gamma}^{(1)}  \,.
\end{equation}
The ``evolution time'' $t$ measures the separation of the scales $\mu_s$ and $\mu_h$: one finds  $t=0$ for $\mu_s=\mu_h$ and a growing $t$ for increasing separation $\mu_s<\mu_h$. As the soft scale approaches the Landau pole, one finds $t\rightarrow\infty$. If the scale  $\mu_h$ is kept fixed the function $t\equiv t(\mu_s)$ is bijective.

The discussion so far applies both to massive and to massless partons. The difference between the two cases becomes evident when one considers the one-loop anomalous dimension matrix
\begin{equation}\label{eq:gammaOne}
\bm{\Gamma}^{(1)} =  \left(
\begin{array}{ccccc}
\, \bm{V}_{2} &   \bm{R}_{2} &  0 & 0 & \hdots \\
0 & \bm{V}_{3} & \bm{R}_{3}  & 0 & \hdots \\
0 &0  &  \bm{V}_{4} &  \bm{R}_{4} &   \hdots \\
0& 0& 0 &  \bm{V}_{5} & \hdots
\\
\vdots & \vdots & \vdots & \vdots &
\ddots \\
\end{array}
\right),
\end{equation}
where the matrix elements $\bm{R}_m$ and $\bm{V}_m$ (which are themselves matrices in color space) are  associated with the emission of a real or virtual soft gluon
\begin{align}
\bm{V}_m  &= 2\,\sum_{i,j=1}^m\,(\bm{T}_{i,L}\cdot  \bm{T}_{j,L}+\bm{T}_{i,R}\cdot  \bm{T}_{j,R})  \int \frac{d\Omega(n_k)}{4\pi}\, W_{ij}^k   \nonumber\\
&\hspace{3cm} - 2\, i \pi \,\sum_{i,j=1}^m \left(\bm{T}_{i,L}\cdot  \bm{T}_{j,L} - \bm{T}_{i,R}\cdot  \bm{T}_{j,R}\right) \Pi_{ij}\, 
, \label{eq:oneLoopRGvirt}\\
\bm{R}_m & = -4\,\sum_{i,j=1}^m\,\bm{T}_{i,L}\cdot\bm{T}_{j,R}  \,W_{ij}^{m+1}\,  \Theta_{\rm in}(n_{m+1})\label{eq:oneLoopRGreal}\,.
\end{align}
The color matrices $\bm{T}_{i,L}$ act on the hard function from the left, i.e. on the amplitude, while $\bm{T}_{i,R}$ act on the conjugate amplitude. The function $\Theta_{\rm in}$ enforces that the hard emission is inside the allowed region. The factor $\Pi_{ij}$ is equal to $+1$ if $i$ and $j$ are both incoming or outgoing legs, and equal to $0$ otherwise. When considering both massless and massive partons,  the dipole radiator takes one of the following forms: 
\begin{align}
\text{massless:}&\hspace{1.5cm}W_{ij}^k = \frac{n_i \cdot n_j}{(n_i \cdot n_k) (n_k \cdot n_j)}\label{eq:dipMasslessMassless} \, ,\\
\text{mixed:}&\hspace{1.5cm}W_{ij}^k = \frac{v_i \cdot n_j}{(v_i \cdot n_k) (n_k \cdot n_j)}\label{eq:dipMassiveMassless}\, ,\\
\text{massive:}&\hspace{1.5cm}W_{ij}^k = \frac{v_i \cdot v_j}{(v_i \cdot n_k) (n_k \cdot v_j)}\label{eq:dipMassiveMassive}\, .
\end{align}
In the special case of $i=j$ (which can not occur in the mixed case \eqref{eq:dipMassiveMassless}, as it implies that the two legs are the same), the radiator  \eqref{eq:dipMasslessMassless} vanishes for massless legs, as $n_i\cdot n_i=0$, but is non-zero for massive quarks \eqref{eq:dipMassiveMassive}. The different kinematics and the presence of the monopoles distinguish the massive from the massless case. 

\begin{figure}[t]
	\centering
	\includegraphics[width=0.8\textwidth]{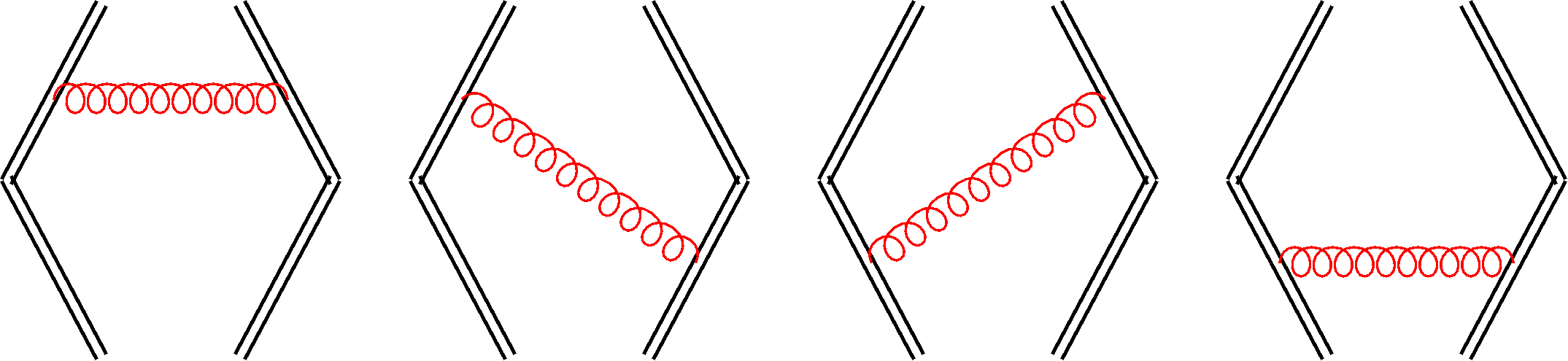}
	\caption{The four possible radiation patterns for a dipole of two massive legs. From left to right: monopole correction to leg 1 corresponding to the term $(i,j) =(1,1)$ in \eqref{eq:oneLoopRGreal}, dipole correction $(1,2)$, dipole correction $(2,1)$, and monopole correction $(2,2)$.   \label{fig:corrDiag}}
\end{figure}

As mentioned above, we work in the large-$N_c$ limit in which the color structure becomes trivial and reduces to factors of $N_c$. This is a huge simplification over the general case in which the $m$-parton terms act in the color-space of the $m$-partons. There is currently a large effort by several groups aiming to extend parton showers beyond the large-$N_c$ case, but we restrict ourselves to this limit. The fact that the color structure becomes trivial implies that the Glauber phases in  $\bm{V}_m$ in \eqref{eq:oneLoopRGvirt} vanish. Furthermore, all interference effects are suppressed and exchanges are only possible between neighbouring legs. However, the monopole contributions are present and need to be included, as is obvious from the diagrams shown in Figure \ref{fig:corrDiag}.

The full corrections in the large-$N_c$ limit read
\begin{align}
\bm{V}_m  &= - 4 N_c \,\bm{1}\,\sum_{i=1}^{m-1} \int \frac{d\Omega(n_k)}{4\pi}\, \widetilde{W}_{i\,i+1}^k  \,, \label{eq:oneLoopRGlargeNcvirt} \\
\bm{R}_m & = 4 N_c \,\bm{1}\,\sum_{i=1}^{m-1}   \widetilde{W}_{i\,i+1}^k \Theta_{\rm in}(n_{k}) \label{eq:oneLoopRGlargeNcreal}\,.
\end{align}
The sum includes all dipoles $i$ consisting of the legs $i$ and $i+1$ and we have absorbed the monopole contributions into the dipoles by defining
\begin{equation}\label{eq:tildeWijk}
\widetilde{W}_{ij}^k \equiv W_{ij}^k -\frac{1}{2}\left(W_{ii}^k+W_{jj}^k\right)\,.
\end{equation}

In the rest of this work, the framework discussed here is applied to top-pair production. In this case the massive legs are always chosen to be the the first and the last in the list of Wilson-line directions, so that monopole radiation can only occur at $i=1$ and $i=m-1$, as the monopole radiator $W_{ii}^k$ is manifestly zero for the massless gluonic legs in between.

In Figure \ref{fig:corrDiag}, we have depicted all possible real emissions for one dipole of two massive Wilson lines. The relative sign of the dipole and monopole contributions in \eqref{eq:tildeWijk} can be understood intuitively by looking at the figure: the partons in the dipole have opposite charge, in contrast the monopole. The factor of two of the dipole term compared to the monopole ones arises because one has to add the identical contribution of the two dipoles $(ij)$ and $(ji)$.  

The details on how one gets from the RG equation to a parton shower are thoroughly explained in \cite{Balsiger:2018ezi}, but for completeness we briefly review the derivation here. The parton shower is based on the RG equation of the hard function which reads
\begin{align}\label{eq:hrdRG}
	\frac{d}{d\ln\mu}\,\bm{\mathcal{H}}_m(\{\underline{n} \},Q,\mu)  &= - \sum_{l =2}^{m}  \bm{\mathcal{H}}_l(\{\underline{n} \},Q,\mu) \, \bm{\Gamma}^H_{lm}(\{\underline{n}\},Q ,\mu) \, .
\end{align}
By changing variable from the scale $\mu$ to the evolution time $t$ and by making use of the fact that the one-loop anomalous dimension matrix has the simple form \eqref{eq:gammaOne}, the evolution equation at LL accuracy takes the form
\begin{align}\label{eq:diffhrd}
\frac{d}{dt}\,\bm{\mathcal{H}}_m(t)  &=   \bm{\mathcal{H}}_m(t) \,  \bm{V}_m +   \bm{\mathcal{H}}_{m-1}(t) \,  \bm{R}_{m-1} \, . 
\end{align}
 This differential equation \eqref{eq:diffhrd} can also be rewritten as an integral equation:
\begin{equation}\label{eq:MCstep}
\bm{\mathcal{H}}_m(t) = \bm{\mathcal{H}}_m(t_0) \,e^{(t-t_0) \bm{V}_m}
+ \int_{t_0}^{t} dt' \,\bm{\mathcal{H}}_{m-1}(t') \, \bm{R}_{m-1}\, e^{(t-t')  \bm{V}_m}\, .
\end{equation}
Starting from \eqref{eq:MCstep}, one can generate the hard functions in an iterative way as 
\begin{align}\label{eq:iterRG}
\bm{\mathcal{H}}_2(t) &= \bm{\mathcal{H}}_2(0) \,e^{t \bm{V}_2} \, ,\nonumber\\
\bm{\mathcal{H}}_{3}(t) &= \int_{0}^{t} dt' \,\bm{\mathcal{H}}_{2}(t') \, \bm{R}_{2}\, e^{(t-t')  \bm{V}_{3}} \, , \nonumber \\
\bm{\mathcal{H}}_{4}(t) &= \int_{0}^{t} dt' \,\bm{\mathcal{H}}_{3}(t') \, \bm{R}_{3}\, e^{(t-t')  \bm{V}_{4}} \, ,\nonumber \\
\bm{\mathcal{H}}_{5}(t) &= \dots ,
\end{align}
 since $\bm{\mathcal{H}}_{k}(0)=0$ for $k>2$. The cross-section at LL finally reads
\begin{align}\label{eq:sigmaLL}
\sigma_{\rm LL}(Q,Q_0) &\equiv \sigma_{\rm tot} \, R(t) \nonumber \\
&=  \sum_{m=2}^\infty \big\langle  \bm{\mathcal{H}}_m(t) \,\hat{\otimes}\, \bm{1} \big\rangle= \big\langle \bm{\mathcal{H}}_2(t) + \int \frac{d\Omega_3}{4\pi} \bm{\mathcal{H}}_{3}(t) +\int \frac{d\Omega_3}{4\pi}\int \frac{d\Omega_4}{4\pi}  \bm{\mathcal{H}}_{4}(t) + \dots \big\rangle \, .
\end{align}
The iterative structure of \eqref{eq:iterRG} is well suited for implementation into a Monte Carlo code which generates successive emissions and thereby also performs the angular phase-space integrals of \eqref{eq:sigmaLL}. For later convenience, we introduced the quantity $R(t)$ given by the ratio of the resummed cross section with a veto to the inclusive cross section $\sigma_{\rm tot}$. At LL accuracy, one can replace $\sigma_{\rm tot}$ by the Born-level result $\sigma_0$.

The inclusion of the massive Wilson lines into the Monte Carlo code is achieved in a straightforward way. The change compared to the massless case boils down to implementing the angular integrations in \eqref{eq:oneLoopRGlargeNcvirt}, where the modified dipole emitter $\widetilde{W}_{ij}^k$ replaces the massless one. A general algorithm for the evaluation of the angular integrals is discussed in the next section. The details of the Monte Carlo algorithm, which showers tree-level event files obtained by means of  {\sc MadGraph5\Q{_}aMC@NLO}  \cite{Alwall:2014hca}, are presented in Appendix~\ref{sec:MCalg}.

\section{Evaluation of the massive angular phase space integrals} \label{sec:evaluation}

The goal of this section is to evaluate the integral 
\begin{align}
\int\frac{d\Omega(n_k)}{4\pi}\,  \widetilde{W}_{ij}^k=	\int\frac{d\Omega(n_k)}{4\pi}\, \left(W_{ij}^k -\frac{1}{2}\left(W_{ii}^k+W_{jj}^k\right)\right) \, \label{eq:angint}
\end{align}
for arbitrary Wilson lines $u_i$ and $u_j$,  which are either both massless ($u_i=n_i$ and $u_j=n_j$), both massive ($u_i=v_i$ and $u_j=v_j$) or one massive and one massless ($u_i = v_i$ and $u_j = n_j$ or vice versa). 

For the discussion below, it is convenient to normalize all  reference vectors in such a way that the zero component of the four-vector is equal to one, i.e. 
	\begin{equation}
	u_i^\mu \equiv \frac{p_i^\mu}{E_i} \, , \label{eq:vconv}
	\end{equation}
where $E_i$ is the energy component of the vector $p_i^\mu$. With this convention, one finds that $u_i^0=u_j^0=n_k^0=1$ and $u_i\cdot u_i=1-\beta_i^2$ with $\beta_{i}=|\vec{p}_i|/E_i=|\vec{p}_i|/\sqrt{\vec{p}_i^2+m_i^2}$. This differs from the definition adopted in heavy-quark effective theory, where one usually normalizes to the mass, i.e. with our convention \eqref{eq:vconv} $v^2 \neq 1$.

The integral in \eqref{eq:angint} is evaluated by first boosting the vectors into the center-of-mass frame of the dipole and by subsequently changing the angular integration variables to (an appropriate generalization of) rapidity. The reader who is not interested in the technical details of the calculation of \eqref{eq:angint} can skip the following discussion and move directly to Section~\ref{sec:massiveeffects}.

\subsection{Boost to the center-of-mass frame}

In order to calculate the integral in (\ref{eq:angint}), it is convenient to first boost the dipole momenta from the laboratory frame into the center-of-mass frame of the dipole. To construct the relevant boost we use a form of the  Lorentz transformation introduced by Householder \cite{householder}
\begin{align}\label{eq:auxLorentz}
	\Lambda_\nu^\mu(\Delta)=\delta_\nu^\mu -\frac{2}{\Delta^2}\Delta^\mu\Delta_\nu,
\end{align}
where $\Delta_\mu= n_\mu-\tilde{n}_\mu$ is the difference of two light-like vectors $n_\mu$ and $\tilde{n}_\mu$. One immediately verifies that $\Lambda^{\mu}_{\nu} (\Delta) n^\nu = \tilde{n}^\mu$,
so the transformation maps $n_\mu \to  \tilde{n}_\mu$. In addition, it is straightforward to check  that $\Lambda^\mu_\rho \Lambda^\rho_\nu =  \delta^\mu_\nu$ and ${\rm det}(\Lambda)=-1$.

The transformation \eqref{eq:auxLorentz} is easily implemented into a computer code and here we use it to construct a boost of two arbitrary time-like or light-like directions $u_i$ and $u_j$ into a frame where these momenta are back-to-back alongside the $z$-axis. The transformation is carried out in three steps, denoted by $X^\mu_\nu$, $B^\mu_\nu$ and $Z^\mu_\nu$. We denote  lab frame vectors $p_i^\mu$ in the three frames reached by each of the transformations  as 
\begin{align}
	p^\nu \stackrel{X^\mu_\nu}{\longrightarrow}\check{ p }^\mu \stackrel{B^\rho_\mu}{\longrightarrow} \tilde{p}^\rho\stackrel{Z^\sigma_\rho}{\longrightarrow} p^{\prime\,\sigma}. \label{eq:deltatrans}
\end{align}
The transformation $X$ rotates the total dipole three momentum such that it points along the $x$-axis. Then $B$ boosts into the center-of-mass frame and the last step $Z$ rotates the system so that the back-to-back vectors lie along the $z$-axis.

Let us now discuss the three transformations in turn. The sum of the momenta associated to the the two vectors $p_i = E_i \,u_i$  and $p_j = E_j \,u_j$  is 
\begin{align}\label{eq:totMom}
	P \equiv p_i + p_j 
   = E (1,\beta\,  \vec{n}_P) \, .
\end{align}

By using the transformation \eqref{eq:auxLorentz} one can find the rotation to a frame where the spatial component of the light-like  vector $ n_P \equiv (1,\vec{n}_P)$ points along the $x$-axis. This rotation (more precisely a rotation with parity inversion since ${\rm det}(\Lambda)=-1$) is defined as 
\begin{align}
X_\nu^\mu\equiv\Lambda_\nu^\mu\left(\Delta_P\right)\, ,
\end{align}
where 
\begin{align}
\Delta_P \equiv n_P - n_X \, , \qquad \text{and} \qquad n_X \equiv (1,1,0,0) \, .
\end{align}
Consequently, by applying the rotation $X^{\mu}_\nu$ to the total momentum one finds
\begin{align}
\check{P}^\mu =X^{\mu}_\nu P^\nu \, , \qquad \text{with} \qquad \check{P} = E (1, \beta ,0,0) \, .  \label{eq:Xrot}
\end{align}

The Lorentz transformation needed to obtain two  back-to-back vectors $\tilde{u}_i$ and $\tilde{u}_j$ from the original vectors $u_i$ and $u_j$ is now a boost along the $x$-axis.
The corresponding transformation in matrix form is
\begin{align} \label{eq:matB}
B \equiv  \left(
\begin{array}{cccc}
\, \gamma &   -\beta \gamma &  \;0 & \;0\\
-\beta \gamma & \gamma & \;0  & \;0  \\
0 &0  &  \;1 &  \;0 \\
0& 0& \;0 & \;1
\end{array}
\right),
\end{align}
where $\beta$ was introduced in \eqref{eq:totMom} and $\gamma=1/\sqrt{1-\beta^2}$. 
Consequently,  the two vectors
\begin{align}
\tilde{p}_i^{\mu} = B^\mu_\rho X^\rho_\nu p_i^\nu & \,, \qquad \tilde{p}_i=\tilde{E}_i(1,\tilde{\beta}_i\vec{\tilde{n}}_i) \, ,
\nonumber \\
\tilde{p}_j^{\mu}=B^\mu_\rho X^\rho_\nu p_j^\nu 
& \, , \qquad
\tilde{p}_j=\tilde{E}_j(1,\tilde{\beta}_j\vec{\tilde{n}}_j) \, ,
\end{align}
are in a back-to-back configuration, i.e. 
\begin{align}
\tilde{E_i} \tilde{\beta}_i \vec{\tilde{n}}_i = 
-\tilde{E_j} \tilde{\beta}_j \vec{\tilde{n}}_j \, . 
\end{align}
Finally, it is convenient to apply a last rotation in order to align the vectors along the $z$-axis. This can be achieved by employing again a Lorentz transformation of the form described in \eqref{eq:auxLorentz}. In particular one can define 
\begin{align} \label{eq:matZ}
	Z_\nu^\mu \equiv \Lambda_\nu^\mu\left(\Delta_Z\right) \, ,
\end{align}
with
\begin{align}
\Delta_Z \equiv \tilde{n}_i  - \tilde{n}_Z \, , \quad \text{and} \quad 
 \tilde{n}_i \equiv (1,\vec{\tilde{n}}_i) \, , \qquad
 \tilde{n}_Z \equiv (1,0,0,1) \, .
\end{align}
In conclusion, the complete Lorentz transformation of any vector from the lab frame into a frame where  the vectors $u^\prime_i = p^\prime_i/E^\prime_i$  and $u^\prime_j = p^\prime_j/E^\prime_j$ are back to back and aligned along the $z$-axis is 
\begin{align}\label{eq:bstToBtB}
L_\nu^\mu \equiv Z_\rho^\mu\,B_\sigma^\rho\,X_\nu^\sigma \,.
\end{align}
One finds that ${\rm det}(L)=1$, since $L$ is the product of one proper and two improper transformations.

\subsection{Evaluation of the angular integral}

After applying the Lorentz transformation $L$ and normalizing the vectors according to \eqref{eq:vconv}, one finds
\begin{align}
 u^\prime_i &= \left(1,0,0,\beta^\prime_i\right) \, , &  u^\prime_j &=  \left(1,0,0,-\beta^\prime_j\right). 
\end{align}
In this frame, one can write the integral over the dipole as 
\begin{align}
\int\frac{d\Omega(n_k)}{4\pi}\, W_{ij}^k &=\int\frac{d\Omega(n_k)}{4\pi}\, \frac{u_i\cdot u_j}{u_i\cdot n_k \, n_k\cdot u_j}\nonumber\\
&=\frac{1 + \beta^\prime_i \beta^\prime_j}{\beta^\prime_i + \beta^\prime_j }\int_0^{2\pi}\frac{d\phi^\prime}{2\pi}\int_{y_{\rm max}}^{y_{\rm max}} dy^\prime\nonumber\\
&=\frac{1 + \beta^\prime_i \beta^\prime_j}{\beta^\prime_i + \beta^\prime_j }\left(y_{\rm max}-y_{\rm min}\right) \, , 
\label{eq:dipInt}
\end{align}
where the light-like momentum $n_k$ in the center-of-mass system is parameterized as 
\begin{align}
n^\prime_k=\left(1,\sin{\theta}\cos\phi,\sin{\theta}\sin\phi,\cos{\theta}\right) \, ,
\end{align}
and $y$ indicates the rapidity-like quantity
\begin{align}\label{eq:genrap}
	y=\frac{1}{2}\ln\frac{n^{\prime \,0}_k+\beta^\prime_j n^{\prime \,3}_k}{n^{\prime \,0}_k-\beta^\prime_i n^{\prime \,3}_k} = \frac{1}{2}\ln\frac{1+\beta^\prime_j \cos{\theta}}{1-\beta^\prime_i \cos{\theta}} \, .
\end{align}
The boundaries of the rapidity integration are 
\begin{align}\label{boundary}
y_{\rm max} &= \frac{1}{2} \ln \left(\frac{1+\beta^\prime _j}{1-\beta^\prime_i}\right) >0 \, , &  y_{\rm min} &= \frac{1}{2} \ln \left(\frac{1-\beta^\prime _j}{1+\beta^\prime_i}\right) < 0\,.
\end{align}
In the massless limits, $\beta^\prime_i$ and/or $\beta^\prime_j$ become equal to $1$ and $y_{\rm max}$ and/or $y_{\rm min}$ go to infinity. In that case, the collinear divergence in the integral (\ref{eq:dipInt}) needs to be regularized. To this end, we apply a hard cutoff $|y| < y_{\rm cut}$ in numerical computations, in addition to the constraints \eqref{boundary}. We then verify that the physical cross sections are cutoff independent. The specific form of the collinear cutoff we use in our code is given in Appendix \ref{sec:MCalg}, see \eqref{eq:cutoff}. A discussion of different cutoffs can be found in Appendix A of \cite{Balsiger:2018ezi}.

The integral over the monopoles gives
\begin{align}
\int\frac{d\Omega(n_k)}{4\pi}\, W_{ii}^k =\int\frac{d\Omega(n_k)}{4\pi}\, \frac{v_i\cdot v_i}{(v_i\cdot n_k)^2} =\int\frac{d\Omega(n_k)}{4\pi}\, \frac{1-\beta_i^2}{\left(1 - \cos \theta \beta_i\right)^2} =1.
\label{eq:monoInt}
\end{align}

Combining the monopole and dipole contributions, the final result for the virtual correction reads 
\begin{align} 
\bm{V}_m  &= - 4 N_c \,\bm{1}\,\sum_{i=1}^{m} \left[
\frac{1 + \beta^\prime_i \beta^\prime_{i+1}}{\beta^\prime_i + \beta^\prime_{i+1} }\left(y_{\rm max}-y_{\rm min}\right) -\frac{1}{2}\left(\delta_{v_i}+\delta_{v_{i+1}}\right)\right] \, , \label{eq:virtCorr} 
\end{align}
with $\delta_{v_{i}}=1$ if $v_i$ is a time-like direction and zero otherwise. Note that the integration boundaries $y_{\rm max}$ and $y_{\rm min}$ depend on $\beta^\prime_i $ and $\beta^\prime_j$, see \eqref{boundary}.

The real emission corrections
\begin{align}
\bm{R}_m & = 4 N_c \,\bm{1}\,\sum_{i=1}^{m}  W_{i\,i+1}^k \left(1 -\frac{1}{2}\frac{W_{ii}^k+W_{i+1\,i+1}^k}{W_{i\,i+1}^k}\right) \Theta_{\rm in}(n_{k}) \label{eq:realCorr}
\end{align}
are evaluated using Monte Carlo methods by randomly choosing a value of $y'$ and $\phi'$ in the integrand of \eqref{eq:dipInt}. The factor inside the bracket in \eqref{eq:realCorr} is a positive weight factor, as shown below. To see whether a given real-emission vector is inside the jet region, one  transforms the vector $n^\prime_k$ back to the laboratory frame by using the inverse transformation to $L$ given in \eqref{eq:bstToBtB}.

\subsection{Positive definiteness of $\widetilde{W}_{ij}^k$}\label{sec:posDef}
\begin{figure}[t]
	\centering
	\includegraphics[width=0.45\textwidth]{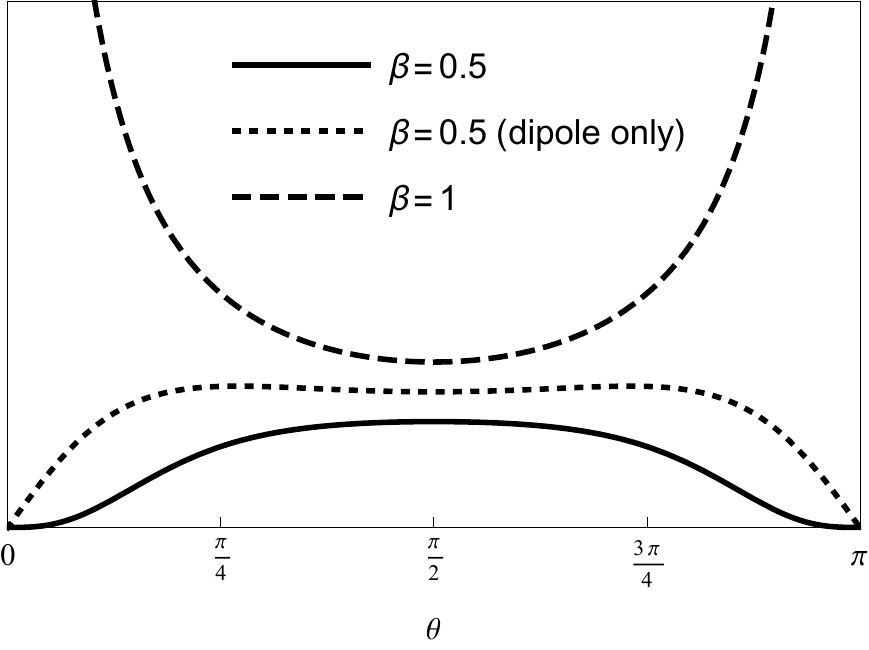}	
	\caption{Angular dependence of the radiation \eqref{eq:realinteg} in the massive case for $\beta_i=\beta_j=0.5$ (solid line) and the massless case $\beta_i=\beta_j=1$ (dashed line). In the massive case, we show the dipole contribution separately (dotted line).\label{fig:dipole}}
\end{figure}

We now show that the weight factor in \eqref{eq:realCorr} is positive. This is done most conveniently in the center-of-mass frame. When written in  terms of scalar products, the factor $\widetilde{W}_{ij}^k$ reads
\begin{align}
\widetilde{W}_{ij}^k =\frac{u_i^\prime \cdot u_{j}^\prime}{(u_i^\prime \cdot n_k^\prime) (n_k^\prime \cdot u_j^\prime)}-\frac{1}{2}\left(\frac{u_i^\prime \cdot u_{i}^\prime}{(u_i^\prime \cdot n_k^\prime)^2}+\frac{u_j^\prime \cdot u_{j}^\prime}{(u_j^\prime \cdot n_k^\prime)^2}\right)\,.
\label{eq:integrand}
\end{align}
To see that this expression is indeed non-negative, one replaces the scalar products by 
\begin{align}
u^\prime_i \cdot u^\prime_{j}&= 1+\beta^\prime_i \beta^\prime_j \, , \nonumber \\
u^\prime_i \cdot n^\prime_{k}&=1-\beta^\prime_i\cos\theta\, , \nonumber \\
u^\prime_j \cdot n^\prime_{k}&=1 + \beta^\prime_j\cos \theta\, . \label{eq:scalprod}
\end{align}
By inserting the relations in (\ref{eq:scalprod}) in (\ref{eq:integrand})
one finds
\begin{align}\label{eq:realinteg}
\widetilde{W}_{ij}^k =
\frac{ (\beta_i^\prime +\beta_j^\prime)^2 \,\sin^2 \theta}{2 (1-\beta^\prime_i \cos \theta)^2 (1+\beta^\prime_j \cos \theta)^2 } \, .
\end{align}
Consequently, the factor $\widetilde{W}_{ij}^k$ in (\ref{eq:integrand}) is always larger than or equal to zero.

\section{Emissions from massive partons and non-global logarithms}\label{sec:massiveeffects}

In Section \ref{sec:factorization} we have shown that in the large-$N_c$ limit the monopole contributions can be absorbed into the dipole terms by replacing the usual dipole emitter $W_{ij}^k$  given in \eqref{eq:radfact} by the modified emitter $\widetilde{W}_{ij}^k$ introduced in \eqref{eq:tildeWijk}. It is interesting to compare the massless and massive cases to illustrate the dead cone effect  \cite{Dokshitzer:1991fc,Dokshitzer:1991fd,Ellis:1991qj,Maltoni:2016ays} mentioned in the introduction. In Figure \ref{fig:dipole} we plot the real-emission integrand \eqref{eq:realinteg} multiplied by the measure $\sin\theta$ as a function of $\theta$. The plot shows the collinear divergences at $\theta=0$ and $\theta=\pi$, which are present in the massless case $\beta_i = \beta_j =1$, while the massive integrand vanishes at the end points. One also observes that the monopole contribution significantly reduces the radiation, compared to the pure dipole contribution shown by the dotted line in the plot.

To see what effect the mass has on the size of non-global corrections, we consider the gap fraction in $e^+e^-$ collisions. To define a gap region, we fix a direction $\vec{n}$ for each event and impose a veto $E_{\rm tot}< Q_0$ on radiation outside a cone around this direction. We then define the rapidity of an emission with momentum $k$ as
\begin{equation}
y  = \frac{1}{2}  \ln \frac{k^0+ \vec{n}\cdot \vec{k}}{k^0- \vec{n}\cdot \vec{k}}\,.
\end{equation}
An emission is outside the cone, i.e. inside the gap region, if $|y|<y_{\rm max}$ and we define the gap fraction as
\begin{equation}\label{eq:gapfraction}
R\left(Q_0\right)=\frac{\sigma_{\text{veto}}(Q_0)}{\sigma_{\text{tot}}}\, .
\end{equation}
For massless final-state quarks $e^+e^- \to q \bar{q}$, one has to ensure that the reference vector $\vec{n}$ is chosen such that radiation collinear to the original partons is included to obtain a collinear safe cross section. To do so, one uses for $\vec{n}$ the thrust axis or the direction obtained from running a jet algorithm on the events. In the massive case $e^+e^- \to t \bar{t}$, on the other hand, we are completely free to choose the reference vector and we compare results obtained when choosing $\vec{n}$ collinear or perpendicular to the top-quark direction. 

\begin{figure}[t]
	\centering
	\begin{tabular}{ccc}
	\includegraphics[width=0.45\textwidth]{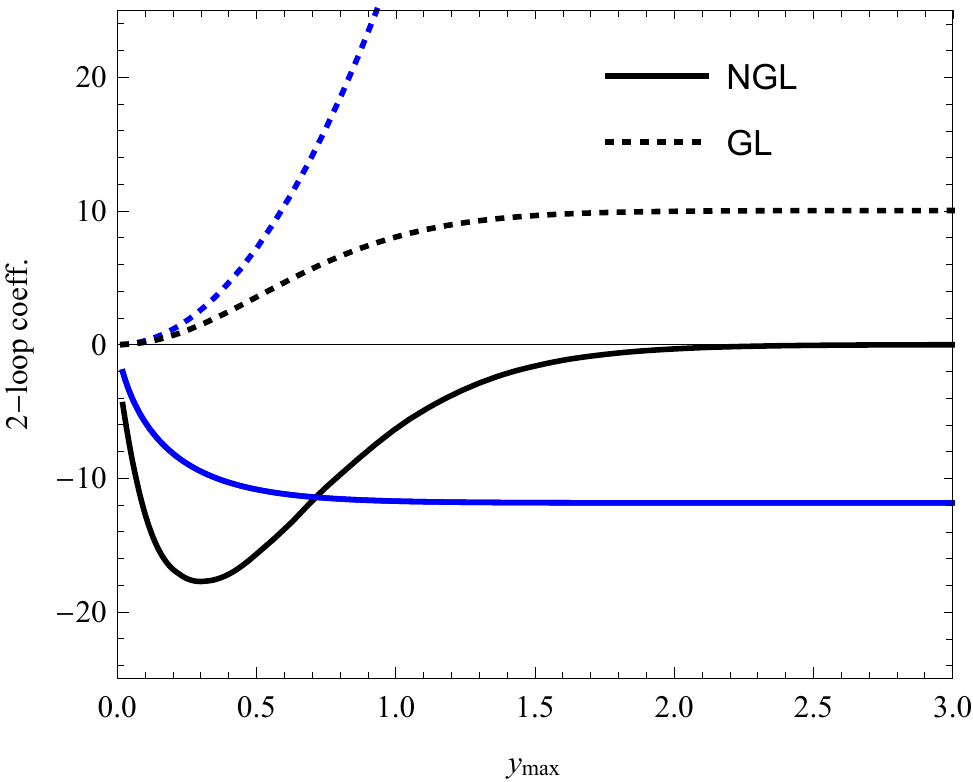}	&& \includegraphics[width=0.45\textwidth]{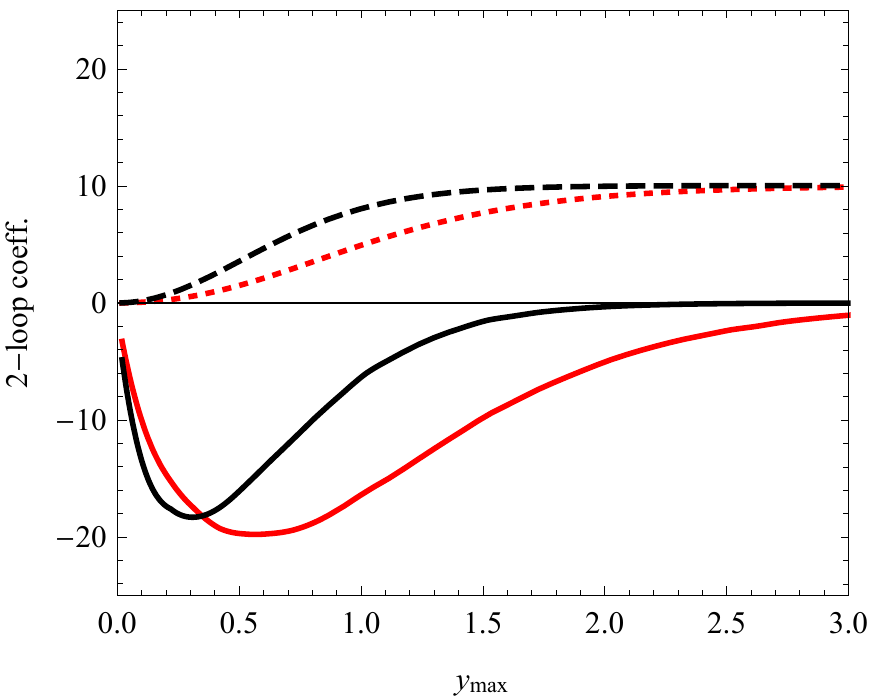}
	\end{tabular}
	\caption{Size of the two-loop terms in \eqref{eq:fixedordercoeff} as a function of the rapidity $y_{\rm max}$ of the gap region. The global contributions $\mathcal{S}_{\rm GL} ^{(2)}$ are shown with dashed lines, the non-global parts $\mathcal{S}_{\rm NGL} ^{(2)}$ using solid lines. The black lines in both panels are identical and correspond to radiation from a massive dipole with  $\beta_i = \beta_j=0.5$ and a reference vector $\vec{n}$ along the direction of the massive quarks. Left panel: Comparison to the massless case (blue lines). Note that the massless coefficients have been divided by 10 to make their size similar to the massive ones.  Right panel: Comparison to the same $\beta_i = \beta_j =0.5$ dipole with $\vec{n}$ perpendicular to the massive quarks (red lines). \label{fig:comparison}}
\end{figure}

 To study the contribution from the first two emissions, we expand
\begin{equation}\label{eq:fixedorderexpansion}
R\left(Q_0\right) = R(t) = 1 + \mathcal{S}^{(1)} t + \mathcal{S}^{(2)} t^2 + \dots
\end{equation}
in the evolution time $t$, which is directly related to $Q_0$, see \eqref{eq:evolutiontime}. The coefficients of the expansion can be obtained by iterating the one-loop anomalous dimension which determines the evolution factor \eqref{eq:evolmat} at LL.  Following the steps outlined in Section 5.2 of \cite{Becher:2016mmh} for the massless case, one finds
\begin{align}
\mathcal{S}^{(1)} =& \left \langle {\bm R}_2 \hat{\otimes} \bm{1} + {\bm V}_2 \right \rangle , \nonumber \\
\mathcal{S}^{(2)} =& \frac{1}{2} \left\langle {\bm R}_2  \hat{\otimes}  \left({\bm R}_3  \hat{\otimes}  \bm{1} +{\bm V}_3  \right)  + {\bm V}_2 \left( {\bm R}_2 \hat{\otimes} \bm{1} + {\bm V}_2\right)\right \rangle .
\label{eq:SRV} 
 \end{align}
 The real-emission parts ${\bm R}_m$ of the anomalous dimension in \eqref{eq:oneLoopRGlargeNcreal} generate an additional parton and the symbol $\hat{\otimes}$ indicates the integral over its direction. The angular brackets denote the normalized color trace, which in the large-$N_c$ limit reduces to the trivial trace $\langle \bm{1} \rangle = 1$. Let us first discuss $\mathcal{S}^{(1)}$. We label the initial hard partons as $1$ and $2$ and the newly emitted gluon as $3$. Then 
 \begin{align}
 \left \langle {\bm R}_2 \hat{\otimes} \bm{1}  \right  \rangle=& 4 N_c \int_\Omega  {\bm 3}_{\rm in}  \widetilde{W}_{12}^3\,,
 \end{align}
 where we introduced the short-hand notation
 \begin{align}\label{eq:fixedordercoeffNLO}
\int_\Omega {\bm 3}_{\rm in} = \int \frac{d \Omega (n_3)}{4 \pi} \Theta_{\rm in} (n_3)\,.
 \end{align}
 The virtual correction ${\bm V}_m$ given in \eqref{eq:oneLoopRGlargeNcvirt} has opposite sign and includes an integral over the entire solid angle. Combining it with the real-emission part, one finds that
 \begin{align}
\mathcal{S}^{(1)} =& \left \langle {\bm R}_2 \hat{\otimes} \bm{1} + {\bm V}_2 \right \rangle = - 4 N_c \int_\Omega  {\bm 3}_{\rm out} \, \widetilde{W}_{12}^3  \, , 
\label{eq:Sone} 
 \end{align}
 where ${\bm 3}_{\rm out}= 1- {\bm 3}_{\rm in}$. 
 The dipole structure after the first emission is $(\bar{q},g,q) = (1,3,2)$. To be consistent with the notation in the anomalous dimensions \eqref{eq:oneLoopRGlargeNcvirt}  and \eqref{eq:oneLoopRGlargeNcreal} one should relabel the particles after the emission as $(1,2,3)$, but we prefer to keep the original labels so that the neighboring dipoles in the second step are $(1,3)$ and $(3,2)$, and
\begin{align}
\left\langle {\bm R_2}\hat{\otimes}  {\bm R_3}\hat{\otimes} {\bm 1} \right\rangle   =&(4 N_c)^2 \int_\Omega  {\bm 3}_{\rm in}\, {\bm 4}_{\rm in}\, \widetilde{W}_{12}^3 \,\left(\widetilde{W}_{13}^4 + \widetilde{W}_{32}^4 \right)   \, .
\end{align} 
We can rewrite all terms appearing in \eqref{eq:SRV} in terms of angular integrals and combine real and virtual parts as we did in \eqref{eq:Sone}. This leads to the two-loop result
\begin{align}\label{eq:fixedordercoeff}
\mathcal{S}^{(2)} &= \mathcal{S}_{\rm NGL} ^{(2)} + \mathcal{S}_{\rm GL} ^{(2)} \nno \\
&=  \, \frac{(4N_c)^2}{2!}  \int_{\Omega}\Big[ -{\bm 3}_{\rm in}\,{\bm 4}_{\rm out}  \widetilde{W}_{12}^3 \left(\,\widetilde{W}_{13}^4+ \widetilde{W}_{23}^4- \widetilde{W}_{12}^4\right) + {\bm 3}_{\rm Out}\,{\bm 4}_{\rm Out}\,\widetilde{W}_{12}^3\,\widetilde{W}_{12}^4\Big]\, ,
\end{align}
in agreement with the results given in \cite{Becher:2016mmh} for the massless case.
The global part of $\mathcal{S}^{(2)} $ is just one half of the squared one-loop contribution, while the non-global piece has a more complicated structure that arises from the emission of a second gluon from the one produced in the first emission. 

Figure \ref{fig:comparison} shows the two-loop coefficients in different situations. In the left plot, where the cone is chosen along the direction of the original dipole, we compare the massive case with $\beta=1/2$ to the massless one. In the massless case (shown in blue) $\mathcal{S}^{(1)} \propto y_{\rm max} $ so that the global part increases quadratically as $y_{\rm max}$ is increased. In the massive case (shown in black), on the other hand, the radiation stops as the gluon becomes collinear to the quark so that the global part of the gap fraction goes to a constant as $y_{\rm max}$ becomes large. Interestingly, the non-global part becomes constant as $y_{\rm max}\to \infty$ in the massless case, while it vanishes for a non-zero mass. The radiation from a massless dipole is much larger than the one in the massive case; indeed it was necessary to divide the massless two-loop coefficients by a factor of ten to make them similar in size to the massive ones in the figure. In the right panel of Figure \ref{fig:comparison} we check how much of a difference the choice of the cone vector $\vec{n}$ makes. The red curves show the result when $\vec{n}$ is chosen perpendicular to the direction of the massive quarks instead of collinear to them. In this case, the massive quarks lie in the middle of the gap region. We observe that the size of the two-loop coefficients for the two choices of $\vec{n}$ is quite similar.

\begin{figure}[t]
	\centering
	\begin{tabular}{ccc}
		\includegraphics[width=0.45\textwidth]{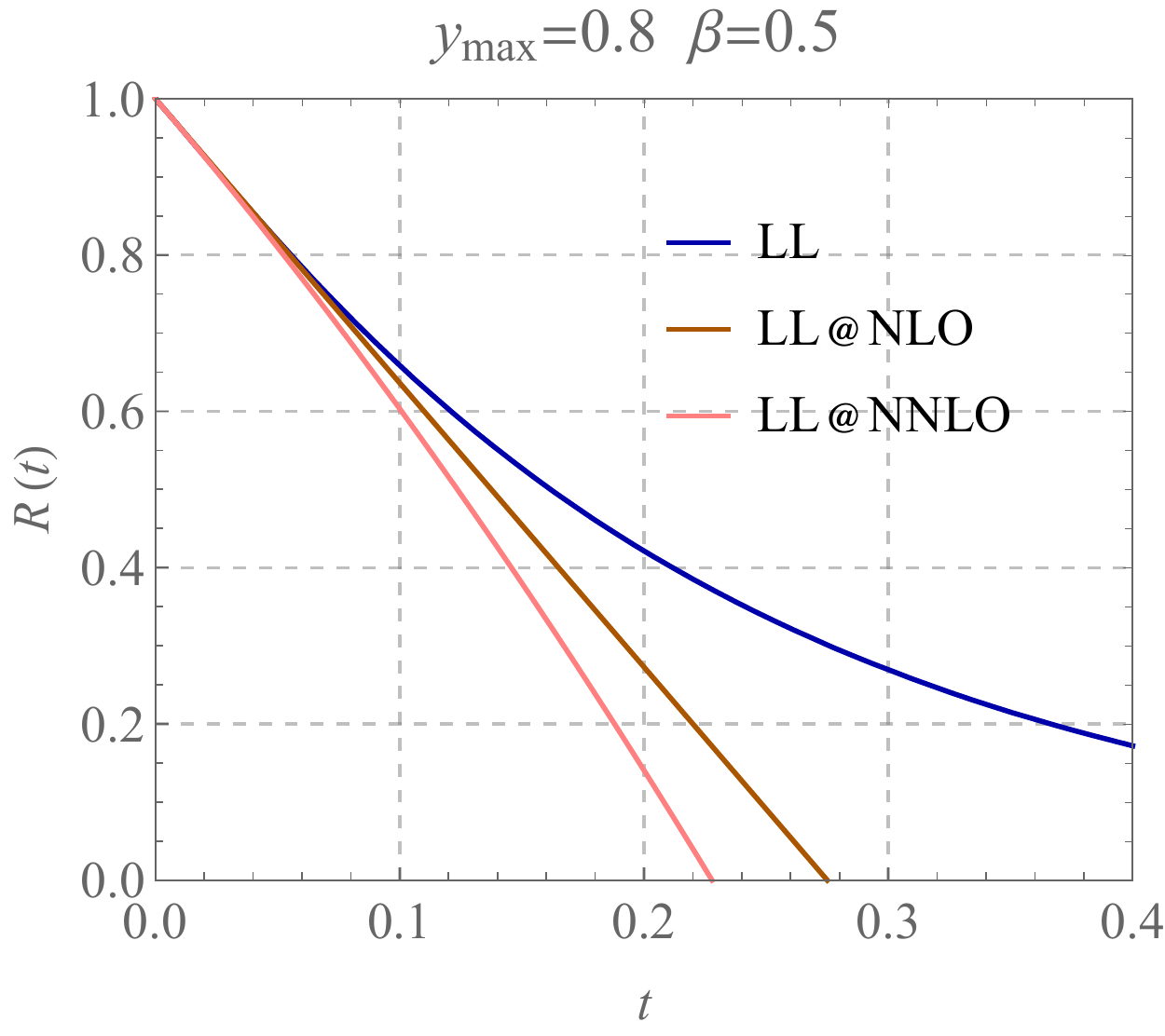}	&& \includegraphics[width=0.45\textwidth]{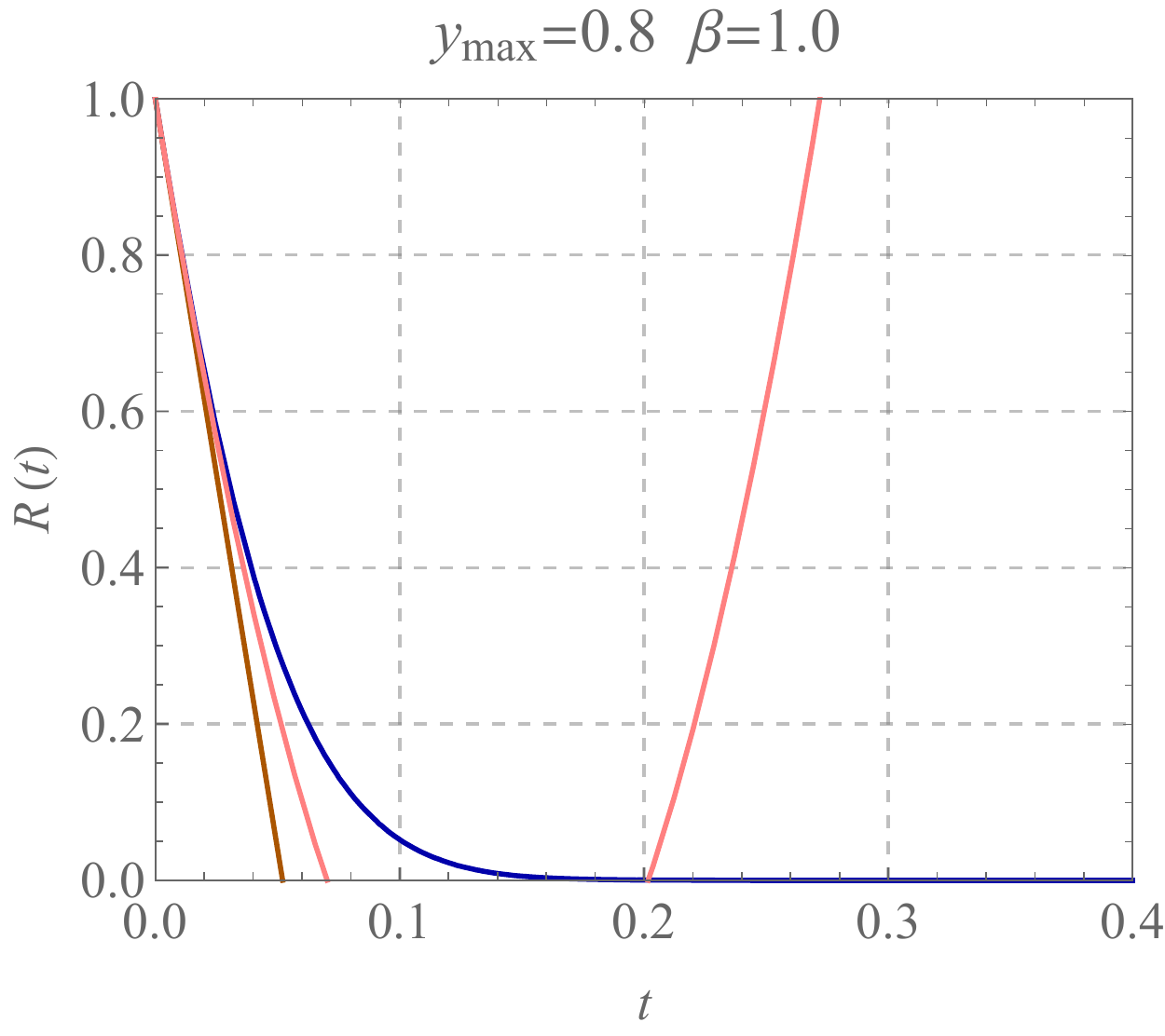}
	\end{tabular}
	\caption{Results of the LL resummation of the gap fraction from a dipole along the cone direction $\vec{n}$. The left plot shows a massive dipole with $\beta=0.5$, the right one a massless one. The LL resummed result is shown in blue,  fixed-order expansions at NLO in brown and at NNLO in pink. \label{fig:dipolesResummation}}
\end{figure}

Having discussed the two-loop corrections, it is interesting to see how the fixed-order expansions compare to the full resummed result. In Figure \ref{fig:dipolesResummation} we show the result of the LL resummation of the gap fraction starting with a single dipole in the center of mass along the cone axis $\vec{n}$ for a gap with maximal rapidity $y_{\rm max}=0.8$. The left plot shows the result for a massive dipole ($\beta=1/2$) while the one on the right starts with a massless one ($\beta=1$). Along with the full LL result, we also plot its NLO and NNLO expansion.  The point made above is fully confirmed; the radiation from a massless dipole is much stronger than from a massive dipole. In fact, both the one-loop and two-loop coefficients are an order of magnitude larger for the massless case than for the massive one. The figure shows the gap fraction as a function of $t$. The relation among $t$ and $Q_0$ depends on the value of $Q$. We stress that the larger values of $t$ in the figure correspond to very small values of $Q_0$. Indeed, for $Q_1= 1\,{\rm TeV}$, $t \geq 0.1$ corresponds to $Q_0 \lesssim 1\,{\rm GeV}$.

\section{\boldmath Resummation of $t\bar{t}$ production with veto on central jets \unboldmath} \label{sec:resummation}

In this section, the formalism is applied to the resummation of non-global logarithms in a cross section involving soft radiation from top quarks. We consider $t\bar{t}$ production at the LHC  with a veto on additional central jet activity as measured by ATLAS \cite{ATLAS:2012al}. This measurement was performed to test the modeling of soft radiation from top quarks in parton shower Monte Carlo codes and is therefore well suited to study resummation effects.

In the measurement ATLAS considers events with at least two energetic $b$-jets, opposite-sign leptons and missing energy, subject to a set of selection requirements designed to enhance the $t\bar{t}$ signal and reject background. In detail, the imposed cuts are as follows: Two of the $b$-jets must have $p_T>25\, {\rm GeV}$, $|y|<2.4$ and $\Delta R(j, l) > 0.4$, where $\Delta R (x,y)=\sqrt{(\Delta \phi(x,y))^2+(\Delta \eta(x,y))^2}$ with  $\Delta \phi (x,y)$ and $\Delta \eta (x,y)$ being the difference of the azimuthal angle and the rapidity of particles $x$ and $y$. 
The opposite charged leptons must fulfill the usual ATLAS cuts: for muons $p_T>20\,{\rm GeV}$, $|\eta|< 2.5$ and for electrons $p_T>25\,{\rm GeV}$, $|\eta|< 2.47$. If the two leptons are of the same flavor, one imposes that their invariant mass is not too small, $m_{\ell\ell} >15\,{\rm GeV}$, and not near the $Z$-resonance, $|m_{\ell\ell}-m_Z| >10\,{\rm GeV}$. In addition one requires missing $E_T^{\rm miss} > 40\,{\rm GeV}$. In the mixed-flavor $\mu e$-channel, one instead imposes that $H_T> 130\,{\rm GeV} $, where $H_T$ is the scalar sum of the visible transverse momenta.

Starting with this event sample, ATLAS then defines a gap region as depicted in Figure~\ref{fig:outsideRegion}. The gap consists of rapidity intervals $y_{\rm min} <|y|<y_{\rm max} $, but the bottom-tagged jets are removed from the gap region. In \cite{ATLAS:2012al}, four rapidity regions with various $y_{\rm min}$ and $y_{\rm max}$ are measured. We will focus on the two regions with gap regions $|y|<0.8$ and $|y|<2.1$.
	
For a given region, the gap fraction is defined as the fraction of events which do {\em not} involve a jet with transverse momentum above $Q_0$ in the gap. The luminosity drops out in the ratio so that the gap fraction is the ratio of the corresponding cross sections, which are both computed in the presence of the selection cuts discussed above, as defined in \eqref{eq:gapfraction}.

For our fixed-order predictions we use {\sc MadGraph5\Q{_}aMC@NLO}  \cite{Alwall:2014hca} and the Les-Houches Event (LHE) files produced by this code are taken as an input for our resummation code. We use  NNPDF2.3  leading-order PDF sets, with $\alpha_s(M_Z)=0.130$ \cite{Ball:2012cx}. For the fixed-order prediction of the gap fraction, we use the relation
\begin{equation}\label{eq:fixed}
R\left(Q_0\right) = 1-  \frac{1}{\sigma_{\rm tot}} \int_{Q_0}^\infty dQ_0'\frac{d\sigma}{dQ_0'}\,.
\end{equation}
Up to corrections of $\mathcal{O}(\alpha_s^2)$, we can use lowest-order cross sections in this formula. We obtain these by generating tree-level events for $t\bar{t}$ and $t\bar{t} g$ with {\sc MadGraph5\Q{_}aMC@NLO}. To be able to impose the selection cuts, specifically the exclusion of the bottom-tagged jets, we let the $t\bar{t}$ pair decay into leptons and a $b\bar{b}$ pair. The $b$ and $\bar{b}$ are then acting as centers of a jet with size $R=0.4$ in the plane of azimuthal angle and rapidity. Therefore, a particle $q$ belongs to the gap region, if 
\begin{align} 
\Delta R (b,q)&>0.4,  & \Delta R (\bar{b},q)&>0.4,
&y_{\rm min} &<\left|y(q)\right|<y_{\rm max}  \,.\label{eq:outReg} 
\end{align}

In the plots of this section the fixed-order predictions for the cross section are shown in green. As in any multi-scale problem, it is not clear what default value one should use for the renormalization and factorization scales. The average partonic center-of-mass energy  $\sqrt{\hat{s}}$ for tree-level $t\bar{t}$ events at the LHC with $\sqrt{s}=7\, {\rm TeV}$ in the presence of the ATLAS selection cuts is about $520\,{\rm GeV}$, which is about three times bigger than the top-quark mass and significantly larger than $Q_0$, the lowest scale in the problem.    We use an intermediate value $\mu_r=\mu_f=2 m_t$ as the  default choice for the renormalization and factorization scales, but one could argue that the relevant scale for $\alpha_s$ in the ratio in \eqref{eq:fixed} is a lower value $\mu_r \sim \mu_s \sim Q_0$ since the factors of $\alpha_s$ associated with the production cross section drop out in the ratio and only the coupling constant associated with the soft gluon emission remains. Indeed, choosing a lower scale would somewhat improve the agreement of the fixed-order prediction with data. The fixed-order uncertainty bands in the plots come from varying the scales $\mu_r$ and $\mu_f$ by factors of two around their default values while imposing $1/2 \leq \mu_r/\mu_f \leq 2$, i.e. we are using the $7$-point method to get the scale bands.  Looking at the predictions for different scale values, we observe that the largest variations arise when both scales are simultaneously varied up or down. The fixed-order scale bands are fairly narrow. While the cross sections themselves have a relatively large scale uncertainty, most of it drops out in the ratio in \eqref{eq:fixed}.

Let us now turn to the resummation, which is performed on the basis of the LHE files for $t\bar{t}$ production. The shower code reads out the momenta of the top quarks and the initial-state particles in order to obtain the directions of the initial Wilson lines. These, together with the large-$N_c$ color dipole structure provided in the event file are the starting point for the shower, which then emits gluons until an emission goes into the gap. The value of the evolution time $t\equiv t(\mu_h,\mu_s)$ in \eqref{eq:evolutiontime}  is later translated into a value of $Q_0$, the scale associated with the emission. The shower also calculates the angular integrals in \eqref{eq:sigmaLL} as it evolves from the hard to the soft scale.

As the default hard scale we use $\mu_h=Q_1=150\,{\rm GeV}$ which was calculated using \eqref{eq:hardest} with an average $Q=\sqrt{\hat{s}}\approx520\,{\rm GeV}$. The soft scale $\mu_s$ should be chosen to be of the order of $Q_0$. However, we want to switch off the resummation at larger $Q_0$ values where we enter the fixed-order regime. To this end, we use a profile function which switches off the resummation for $Q_0 \to Q_{\rm max}$. We choose $Q_{\rm max} = \mu_h = 150\,{\rm GeV}$ and use the same functional form as in \cite{Balsiger:2019tne}, namely
\begin{align}
\mu_{s}=\frac{x_s Q_0}{1 + \frac{x_s Q_0}{\mu_h} - 4 \hat{Q}_0 +6 \hat{Q}_0^2  - 4 \hat{Q}_0^3 +\hat{Q}_0^4}\, , \label{eq:profFun}
\end{align}
where $\hat{Q}_0 = Q_0/Q_{\rm max}$. The profile function is constructed such  that $\mu_s\to x_s Q_0$ for $Q_0 \to 0$ and that $\mu_s\to \mu_h$ for $\hat{Q}_0 \to 1$. The higher-power terms in the denominator are chosen such that the first few derivatives at $\hat{Q}_0 = 1$ vanish and the parameter $x_s=\{\frac{1}{2},1,2\}$ is used for scale variation. Beyond $Q_0=Q_{\rm max}$ all resummation effects are switched off and only the fixed-order prediction remains.

\begin{figure}[t]
	\centering
	\includegraphics[width=0.45\textwidth]{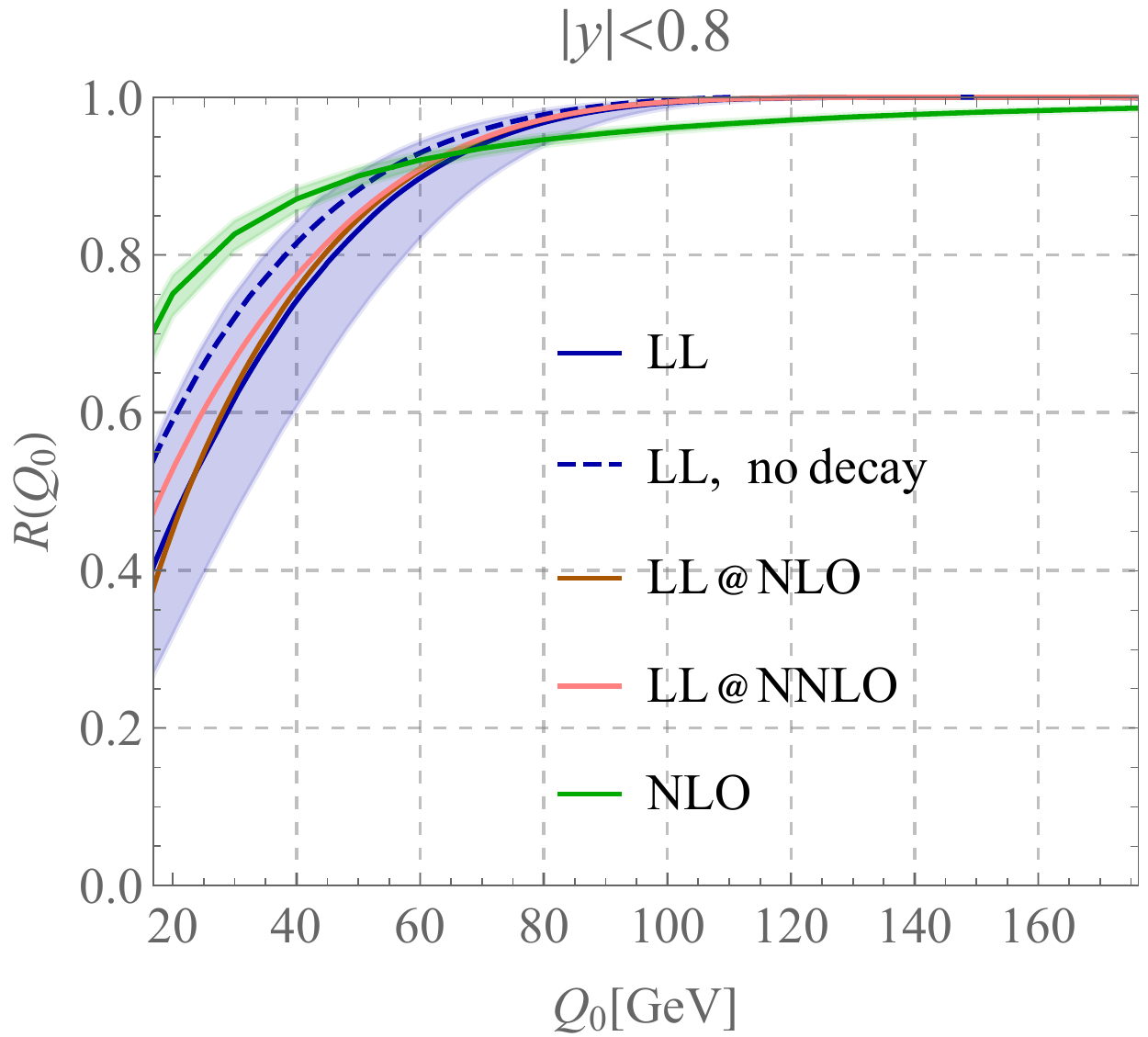}\includegraphics[width=0.45\textwidth]{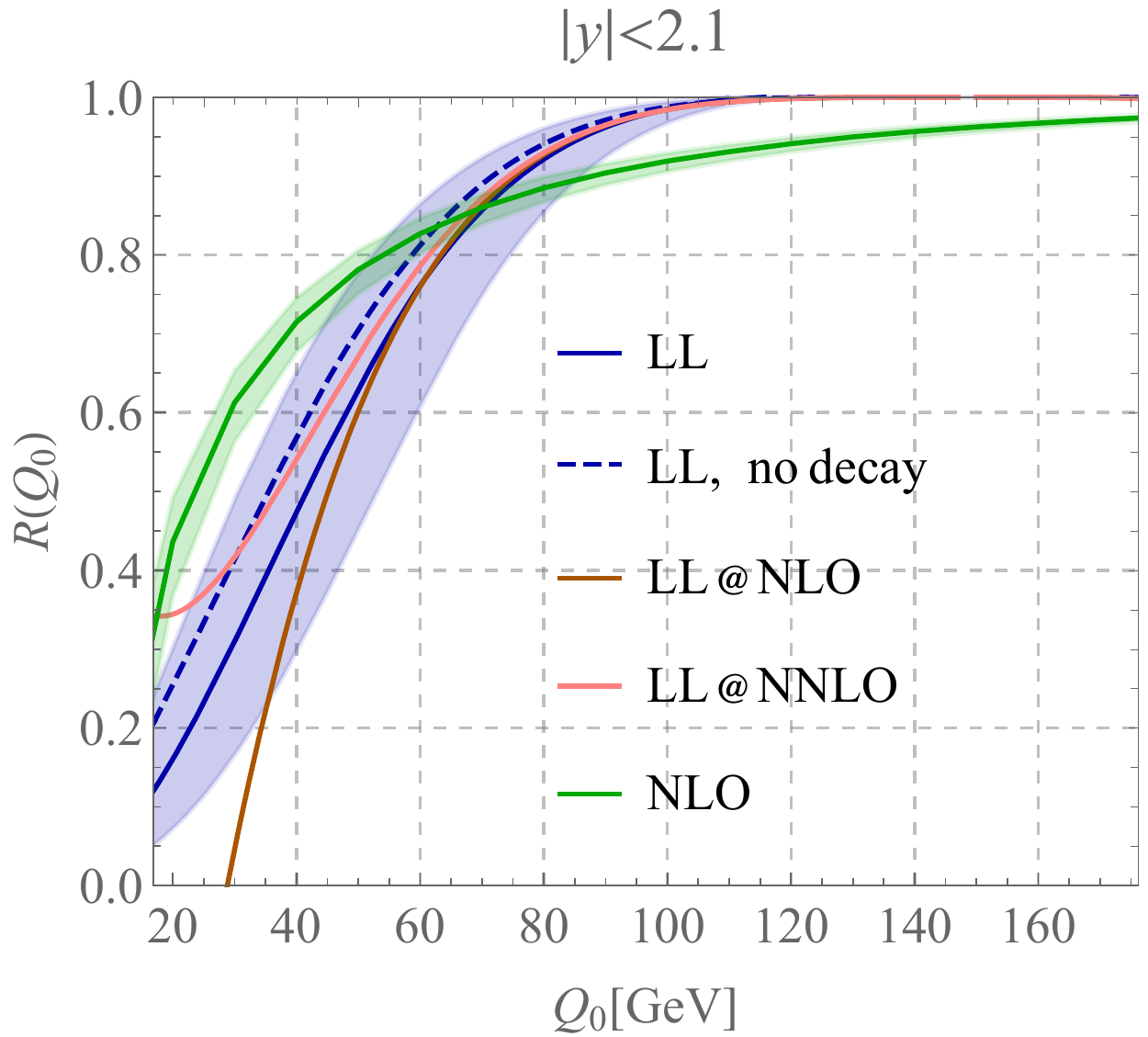}
	\caption{Results of the resummation of the non-global logarithms in $t\bar{t}$ production at $\sqrt{s} = 7~\text{TeV}$ with a veto on additional jets in the two regions with $|y| < 0.8$ (left) and $|y| < 2.1$ (right). 
			The full leading logarithmic resummed result is shown in  blue, its expansions to NLO in brown and to NNLO in pink. The blue dashed line is the  resummed result when the radiation from the decay is omitted. The fixed-order calculation to NLO is shown in green. The uncertainty bands are from scale variation, see text. \label{fig:resummationResults}}
	\end{figure}

The results of the resummation (blue) along with its fixed-order expansion to the second order (brown and pink) as well as the fixed-order calculation results (green) are given in Figure \ref{fig:resummationResults}. From this plot, one can clearly see that the difference between the LL result (blue) and its first-order expansion (brown) is moderate for the gap region with  $|y|<0.8$, while this difference is very large when the gap covers the interval $|y|<2.1$. The effect of the radiation from the top decays is not negligible and reduces the gap fraction. By leaving it out (dashed blue line) one obtains a gap fraction that is sizeably larger especially at low $Q_0$.

The ultimate goal of this section is to match the NLO calculation to the LL resummation in order to obtain LL$+$NLO predictions.
The size of the difference between the LL result and its first order expansion, which we denote by LL@NLO, is relevant for the matching procedure, as discussed below. 

When expanding the LL resummed result, we also expand the evolution time
\begin{equation}
t = \frac{\alpha_s(\mu_r)}{4\pi} \ln \frac{\mu_h}{\mu_s} -\left(\frac{\alpha_s(\mu_r)}{4 \pi}  \right)^2\beta_0 \ln ^2\frac{\mu_h}{\mu_s} +\mathcal{O}(\alpha_s^3)\, 
\end{equation}
so that the expanded LL result depends on $\mu_r$, $\mu_h$, $\mu_s$ as well as the factorization scale $\mu_f$ at which the PDFs are evaluated. To estimate the scale uncertainties of the matched result, we vary the scales $\mu_r$, $\mu_h$, $\mu_s$ and $\mu_f$ individually by a factor of two and then take the envelope.  It turns out that the variation of the soft scale is dominant throughout the plot.

\begin{figure}[t]
	\centering
	\includegraphics[width=0.45\textwidth]{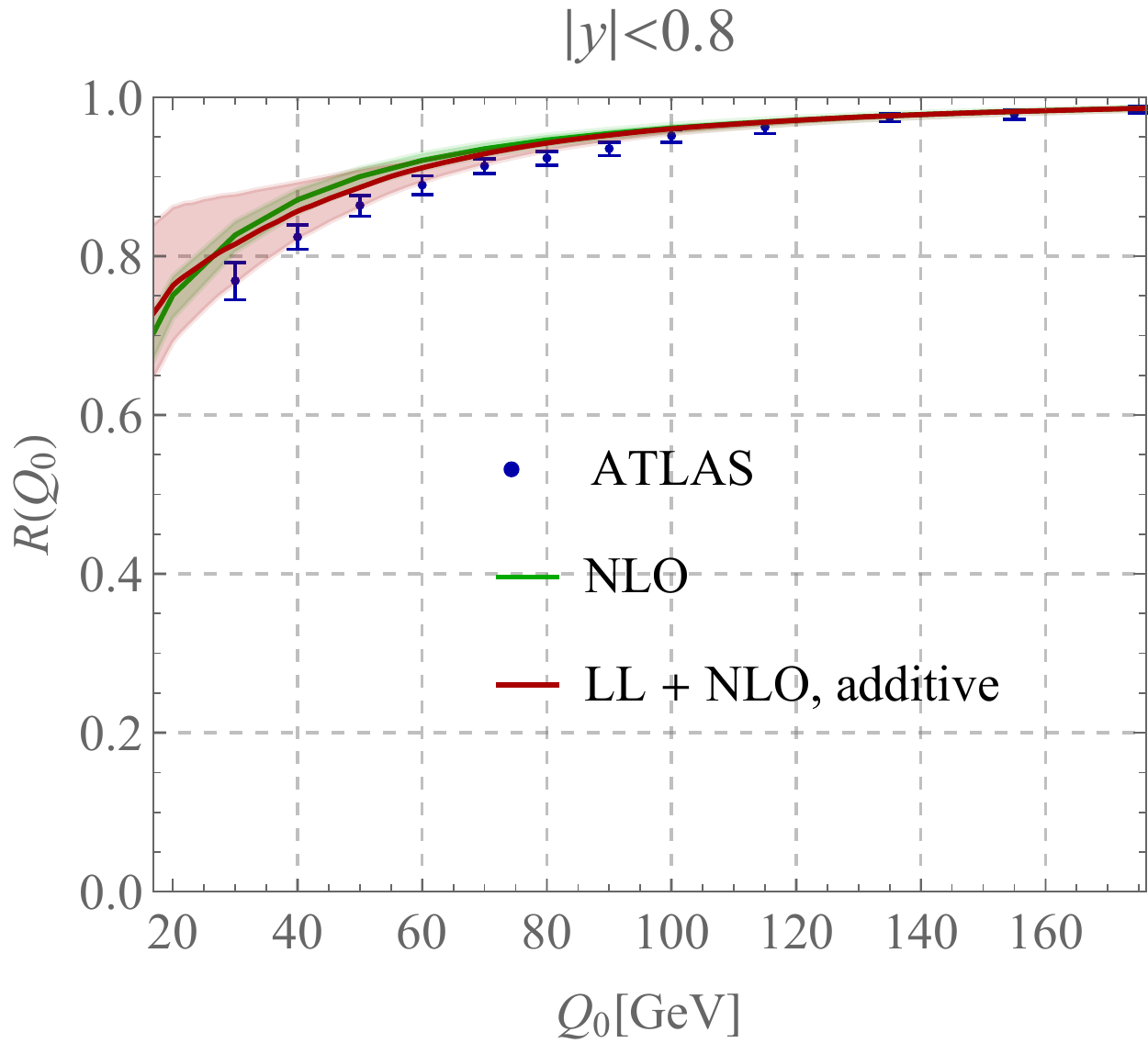}\includegraphics[width=0.45\textwidth]{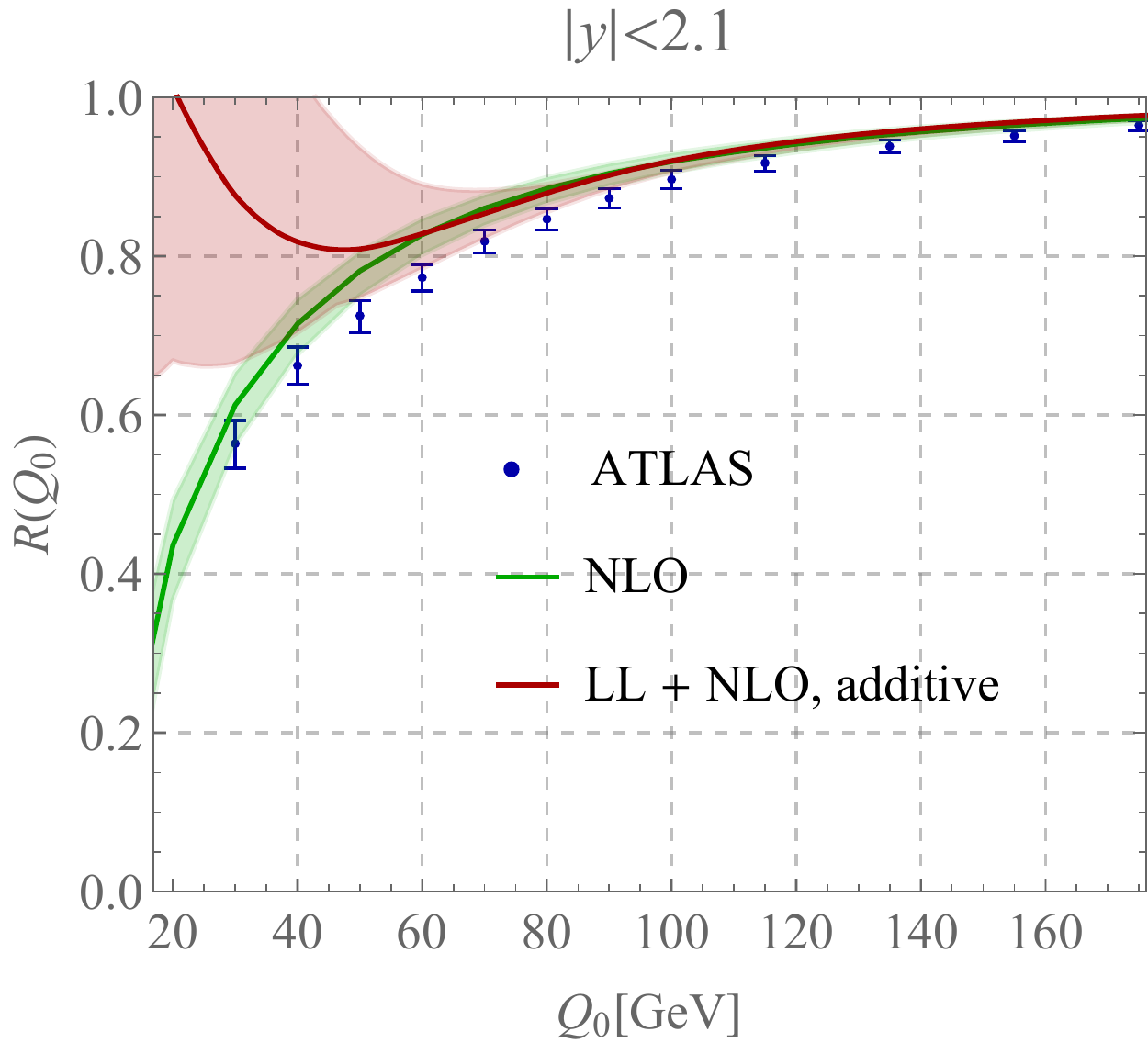}	\caption{Results of the resummation of the non-global logarithms in $t\bar{t}$ production at $\sqrt{s} = 7 \text{TeV}$ with a veto on additional jets in the two regions with $|y| < 0.8$ (left) and $|y| < 2.1$ (right). Shown are the ATLAS measurements (blue points with error bars), the fixed order result (green bands) and the resummed and matched cross section (red bands). The uncertainty bands are from scale variation, see text.\label{fig:matchedResults}}
\end{figure}

There are two common schemes to combine the resummed results and fixed-order predictions, namely  additive and  multiplicative matching. In the additive matching scheme, one simply adds the LL gap fraction to the NLO prediction and subtracts the one-loop expansion of the LL gap fraction to avoid double counting
\begin{align}\label{eq:matching}
	R_{\text{additive}}=R_{\rm LL}(\mu_f,\mu_h,\mu_s)+R_{\rm NLO}(\mu_f,\mu_r)-R_{\rm LL @ NLO}(\mu_f,\mu_r,\mu_h,\mu_s)\, .
\end{align} 
Predictions obtained with the additive matching scheme (red) are shown in Figure \ref{fig:matchedResults} together with the NLO fixed-order results and data from the ATLAS measurement.
	
\begin{figure}[t]
	\centering
	\includegraphics[width=0.45\textwidth]{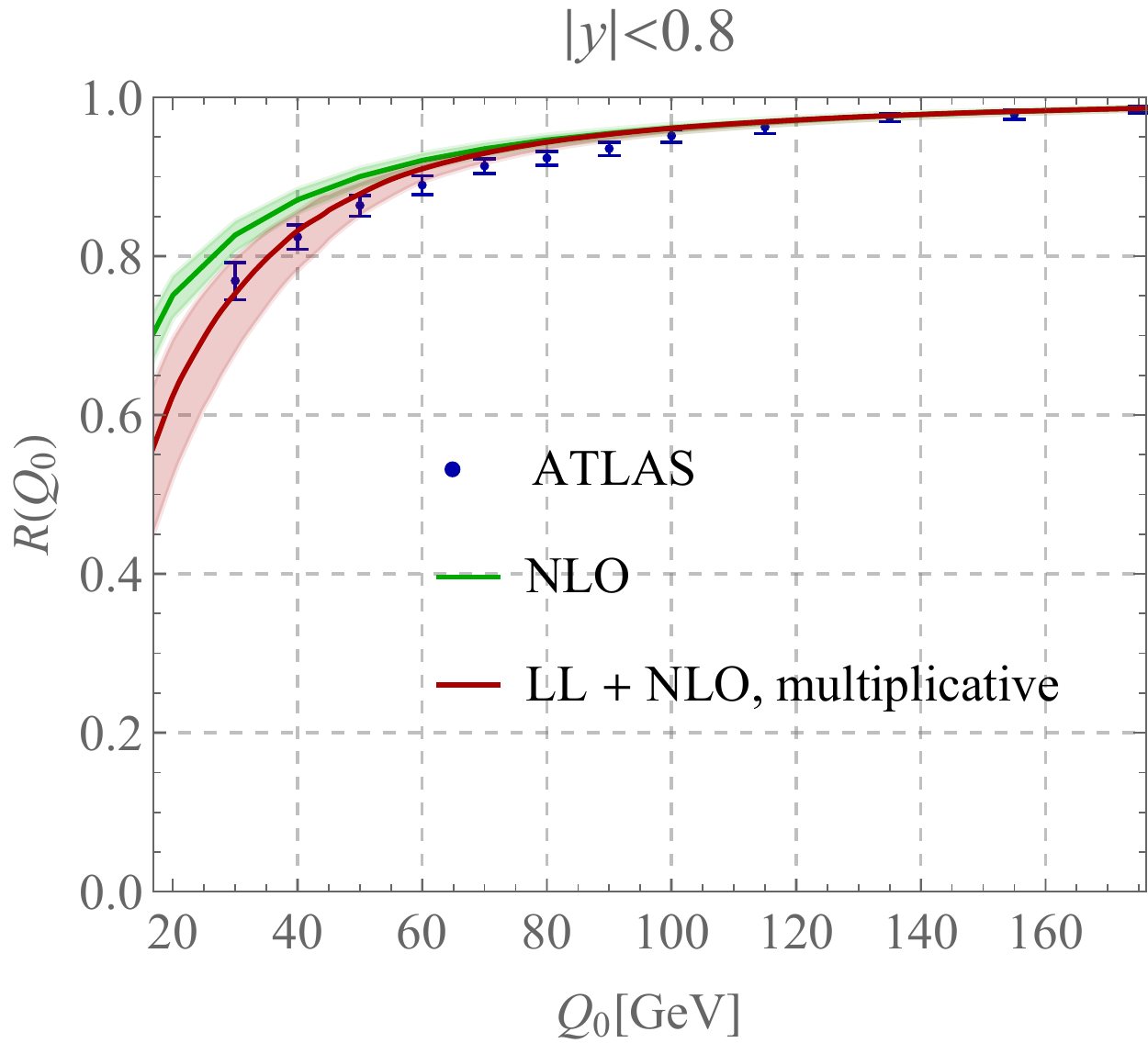}\includegraphics[width=0.45\textwidth]{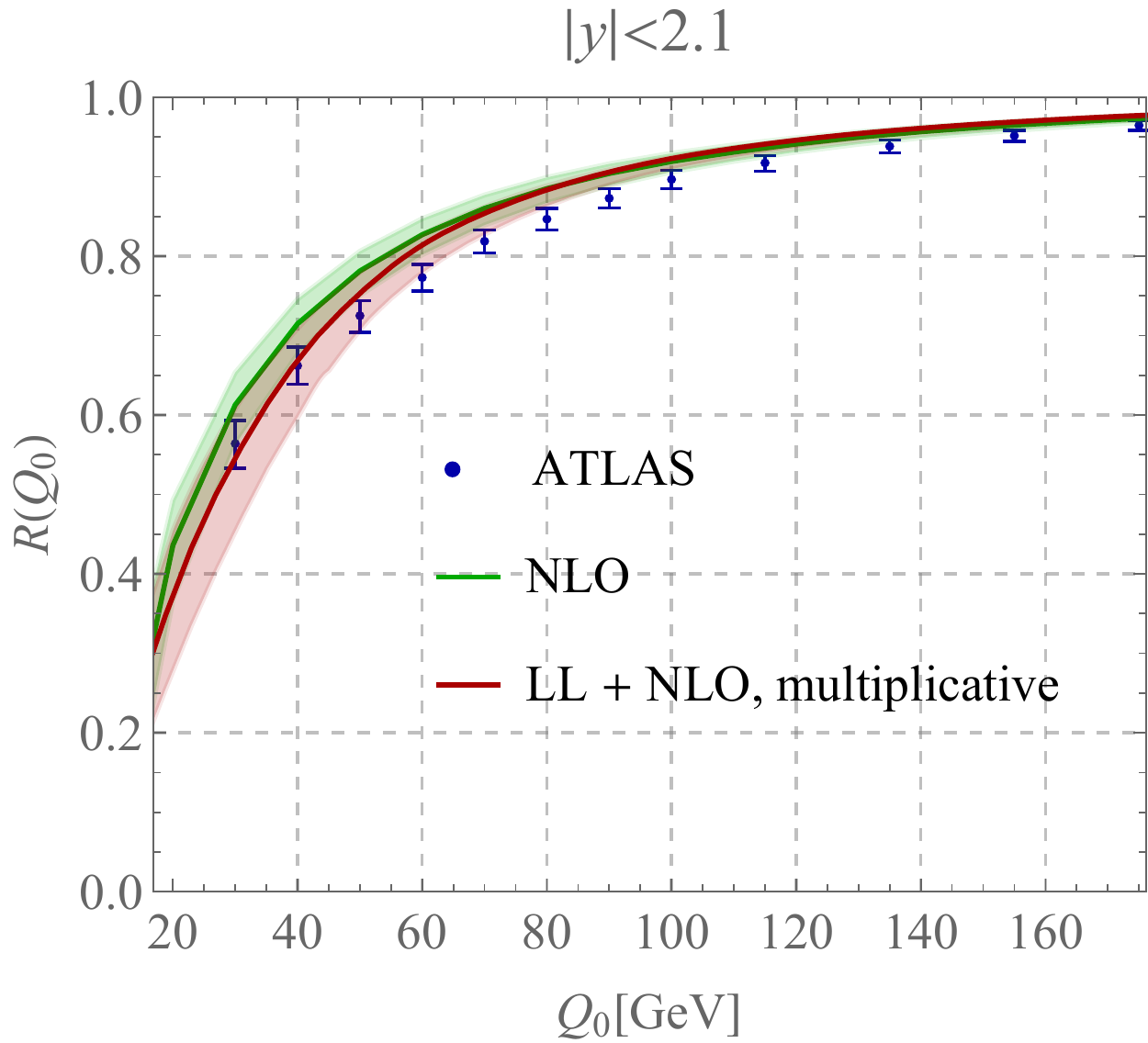}
	\caption{Same as Figure \ref{fig:matchedResults}, except that the multiplicative matching scheme was adopted.\label{fig:matchedResultsMult}}
\end{figure}

In the multiplicative matching scheme, one exponentiates the matching corrections and multiplies the exponential by the LL result
 \begin{align}\label{eq:matchingMult}
R_{\text{multiplicative}}=R_{\rm LL}(\mu_f,\mu_h,\mu_s)\, \exp\!\left( R_{\rm NLO}(\mu_f,\mu_r)-R_{\rm LL @ NLO}(\mu_f,\mu_r,\mu_h,\mu_s)\right)\, .
\end{align}
The results obtained by means of multiplicative matching  are shown in red in Figure \ref{fig:matchedResultsMult}. Multiplicative matching exponentiates the entire first emission, which is similar in spirit to what is done in the {\sc POWHEG} method \cite{Frixione:2007vw,Alioli:2010xd}.  The ATLAS paper compared their measurements to NLO results matched to parton showers using {\sc POWHEG} and also {\sc MC@NLO} \cite{Frixione:2002ik,Frixione:2003ei}. Both schemes reproduce the data to better than 5\%, {\sc POWHEG} is typically even within 1\%-2\% of the measurement. The ATLAS paper does not provide the uncertainties of the theory prediction to which they compare, but we would expect them to be similar in size to the NLO uncertainty bands in our plots. In Section \ref{sec:massiveeffects} it was shown that the radiation from massless legs is numerically much larger than from massive ones. Consequently, we expect that, in order to get a good description of the gap fraction, the modeling of the  initial-state radiation is the most important effect. 
	For this reason, it is not clear to us if a comparison to the ATLAS data provides a sufficiently stringent test of the description of soft radiation from massive quarks in a parton shower.

One observes that the additive matching scheme works well for the gap region $|y|<0.8$ and actually mildly improves the agreement of central value with the data. However, for the case in which the gap region is $|y|<2.1$, the predictions obtained with additive matching become unphysical for small values of $Q_0$. This is not surprising, since the higher-order emissions are enhanced by factors of the gap size $\Delta y$. If these rapidity logarithms become larger, they must be resummed. The formalism to carry out this resummation exists \cite{Becher:2015hka,Becher:2016mmh} but we do not implement it in the present work. 

The multiplicative matching leads to better results since the matched gap fraction correctly vanishes for $Q_0\to 0$, as the resummed result does. Predictions obtained by means of multiplicative matching are shown in Figure \ref{fig:matchedResultsMult}, which shows that they are in good agreement with the experimental data, within the large scale uncertainty bands. To reduce these, it would be important to go to higher logarithmic accuracy, or to at least include higher-order corrections to the hard and soft functions, as it was done in the massless case \cite{Balsiger:2019tne}. 

In order to compare predictions  to the Run I ATLAS measurement  \cite{ATLAS:2012al}, all calculations were carried out at $\sqrt{s}= 7\,{\rm TeV}$. For the tree-level top production process at $\sqrt{s}= 13\,{\rm TeV}$, one finds that the average partonic center-of-mass energy is $Q\approx 550\,{\rm GeV}$, which translates into $Q_1 \approx 170\,{\rm GeV}$, only $20\,{\rm GeV}$ higher than at $7\,{\rm TeV}$. Consequently, we conclude that the result for the gap fraction at $\sqrt{s}= 13\,{\rm TeV}$ would be quite similar to the ones at Run I.

\section{Conclusion }\label{sec:conclusion}

In this paper, we have developed the necessary formalism to carry out the resummation of non-global logarithms for processes involving massive quarks. More specifically, we discussed how the parton shower approach needs to be modified to go beyond the high-energy limit, implemented those changes and then compared the radiation patterns of massive and massless partons. As an application, we have performed the leading logarithmic resummation of the cross section for $t\bar{t}$ production with a veto on central jet activity, an observable measured at  Run I of the LHC. 

Soft radiation has a well-known eikonal form, independent of the mass of the emitting parton. However, in the massive case, the velocity vector of the emitting parton is time-like rather than light-like. This fact makes the kinematics of the process more complicated. A second important difference to the massless case is that for massive emitters one needs to account for monopole radiation. The radiator $W_{ij}^k$ describing an emission between legs $i$ and $j$ is nonzero for $i=j$ for massive legs; therefore the parton shower must also include radiation from a single leg, despite the fact that it is a purely soft shower. We have shown that in the large-$N_c$ limit this radiation can be absorbed into the dipoles by replacing the usual radiator with a modified one, indicated with $\widetilde{W}_{ij}^k$. The monopole radiation has a negative relative sign with respect to the dipole contribution, but the total contribution $\widetilde{W}_{ij}^k$ remains positive. These properties make it straightforward to implement massive partons in the shower code that was previously developed for the emission from massless quarks \cite{Balsiger:2018ezi,Balsiger:2019tne}.

Comparing the two cases, we observe that the massive dipole radiator is numerically significantly smaller than the massless one and that the radiation is further reduced by the monopole terms. For example, when analyzing the fixed-order expansion of the leading logarithmic resummation for a gap in the central rapidity region of size $\Delta y = 1.6$, both the one-loop and the two-loop coefficients are an order of magnitude larger for a massless dipole compared to one with two massive legs with $\beta=0.5$ each. 

ATLAS measured the gap fraction in $t\bar{t}$ production with a veto on central jet activity \cite{ATLAS:2012al}. This provides an interesting test case for the computational framework developed here.  However, to compare to experiment we also need to account for radiation from the top-quark decay. To do so, we work in the narrow-width approximation in which the process factors into production and decay and then apply the parton shower to all color dipoles associated to the $t\bar{t}$ production as well as to the dipoles associated with the decay of the $t\bar{t}$ pair. The predictions that we compare to the ATLAS measurements are obtained by matching the LL resummed result to the NLO fixed-order computation of the gap fraction. There are two schemes commonly used to combine resummed and fixed order results: additive and multiplicative matching. For small gap sizes $\Delta y$ both schemes give similar results while for larger gaps the additive matching  yields unphysical gap fractions. The problems for large gap sizes are not unexpected since the higher order corrections (and also the power-suppressed terms added in the matching) are enhanced by $\Delta y$, i.e.\ by collinear logarithms. If these logarithms become large they must be resummed as well. The formalism necessary to implement this resummation exists \cite{Becher:2015hka,Becher:2016mmh}, but the corresponding  calculation is beyond the scope of the present paper. 

In the present work, we resummed the leading non-global logarithms. In order to go to higher logarithmic accuracy, one needs to include the one-loop corrections to both the hard and the soft function,  the tree-level result of the hard function with one additional emission, and to evolve with the two-loop anomalous dimension matrix. In the massless case,  calculations including the first three ingredients listed above were recently presented \cite{Balsiger:2019tne}. Work on the final ingredient, the two-loop anomalous dimension matrix, is ongoing. 

In this paper, we have extended the resummation of non-global observables to processes involving massive partons in the large-$N_c$ limit. 
Obviously, it would be desirable to extend the formalism to include logarithmic corrections beyond the large-$N_c$ limit.
This would be especially interesting since Glauber phase effects then start to play a role in hadronic collisions. There are a few first finite-$N_c$ results in \cite{Hatta:2013iba,Hagiwara:2015bia} based on a different formalism \cite{Weigert:2003mm} and there is a considerable amount of ongoing work focused on the inclusion of subleading color effects into parton showers \cite{Platzer:2013fha, Platzer:2012np, Nagy:2015hwa, Martinez:2018ffw, Isaacson:2018zdi, Nagy:2019pjp, Forshaw:2019ver, Hoeche:2020nsx, Forshaw:2020wrq}, but a full implementation of all subleading-color effects, in particular Glauber phases, is not yet available.

An understanding of non-global logarithms could prove useful in the context of the top-quark mass determination. Given the complicated structure of these types of logarithms and our limited ability to perform all-order resummations, it is of course desirable to avoid them in the context of precision physics. On the other hand, to maximize sensitivity to the top-quark mass, jet observables are preferable to inclusive cross sections. It has been proposed to use jet substructure techniques such as grooming to reduce the sensitivity to soft radiation \cite{Andreassen:2017ugs} and a factorization theorem implementing grooming has been put forward  \cite{Hoang:2017kmk}. These techniques can reduce the size of non-global logarithms, and our approach could be used to get a better understanding of the remaining effects and their uncertainty.

\begin{acknowledgments}	
The authors thank Paolo Nason and Thomas Rauh for useful comments and discussions. The research of T.B.\ is supported by the Swiss National Science Foundation (SNF) under grant 200020\_182038. The work of A.F.\ is supported in part by the PSC-CUNY Award 62243-00 50. M.B.\ thanks the New York City College of Technology of CUNY for hospitality during a visit July 2019. A.F.\ would like to thank  the Albert Einstein Center for Fundamental Physics at Bern University for hospitality in September and October 2017, during the early stages of this project.
\end{acknowledgments}  

\newpage

\appendix

\section{Details of the Monte Carlo algorithm \label{sec:MCalg}}
This appendix describes in detail the algorithm used to obtain the results presented in Section~\ref{sec:resummation}. We will start with a sample tree-level file and will then show how it is processed step by step by our code. This level of detail is not necessary for most readers, but should be useful for someone implementing a similar shower. It can also serve as a documentation of our code (written in {\sc Python}), which we plan to make public in the future.

 The starting point for the LL resummation algorithm  is a Les Houches Event File (LHEF) \cite{Alwall:2006yp} for the hard process produced using {\sc MadGraph5\Q{_}aMC@NLO} \cite{Alwall:2014hca}. The generated process is the collision of two protons with center-of-mass energy $\sqrt{s}=7\,{\rm TeV}$ producing a $t\bar{t}$ pair. Each top quark in the pair decays in a bottom quark and a $W$-boson. The latter is required to decay  leptonically. In this way, the final state includes  a $b\bar{b}$ pair, two leptons and two neutrinos. In {\sc MadGraph5\Q{_}aMC@NLO} syntax, the process is generated by the following command:
\begin{center}
{\tt	generate p p > t t\textasciitilde\phantom{a}> vl l+ vl\textasciitilde\phantom{a}l- b b\textasciitilde	}
\end{center}
	In $t\bar{t}$ production at leading order, the partonic initial state includes either two incoming gluons (gluon fusion channel) or two incoming quarks (quark annihilation channel). {\sc MadGraph5\Q{_}aMC@NLO} computes the cross section for $N_c=3$ and then randomly assigns one of the possible large-$N_c$ color structures to each tree-level event so that it can be analyzed by a parton shower. The large-$N_c$ color structure is given by a set of dipoles, as illustrated in Figure \ref{fig:dipolestructure}. An event consists of four or five dipoles in total, namely two (quark annihilation) or three (gluon fusion) dipoles associated to the production of the top pair and two dipoles from the radiation in the decay to bottom quarks.

In narrow-width approximation, the amplitudes squared factorize into production and decay. We will separately compute the emissions from production and decay and obtain the cross section as a product
\begin{align}\label{eq:app_multOfDip}
	{\sigma}^{{\rm Event}}_{{\rm LL}}(t)& = \sigma_0^{{\rm Event}} \,R_{t\bar{t}} (t)\, R_{t \to b}(t)\, R_{\bar{t} \to \bar{b}}(t) \, ,
\end{align}
where $\sigma_0^{{\rm Event}}$ is the Born-level event weight supplied by {\sc MadGraph5\Q{_}aMC@NLO}.  The factors $R_{t\bar{t}}(t)$, $R_{t \to b}(t)$ and $R_{\bar{t} \,,\to \bar{b}}(t)$ are computed by showering the color dipoles arising in the production and decay process. The product form \eqref{eq:app_multOfDip} holds on the level of the squared amplitudes in the large-$N_c$ limit, but the observable $Q_0$, the energy inside the veto region, is additive and the cross section will therefore be a convolution of the different pieces, not simply a product. However, at LL accuracy, the cross section is $Q_0$-independent as it only depends on $t\equiv t(\mu_h, \mu_s)$ and the convolution then reduces to the product in \eqref{eq:app_multOfDip}.

Below, we illustrate the parton shower using the production process $R_{t\bar{t}} (t)$, but we run exactly the same shower for the dipoles in the decay.  We consider  an event in the gluon fusion channel to discuss the showering process in detail in the following, since this is the most involved case. The dipole structure of this event is shown in Figure \ref{fig:dipolestructure_App}. We could separately shower each of the three dipoles, but it is more efficient to treat the event as one dipole with two intermediate gluons, see below. The form of the shower for $R_{t \bar{t}}(t)$ is then similar to \eqref{eq:sigmaLL} except that the process starts with four partons
\begin{align}\label{eq:Rttb}
 \sigma_0^{t\bar{t} }\, R_{t \bar{t}}(t)=\mathcal{H}_4(t) + \int \frac{d\Omega_5}{4\pi} \mathcal{H}_{5}(t) +\int \frac{d\Omega_5}{4\pi}\int \frac{d\Omega_6}{4\pi}  \mathcal{H}_{6}(t) + \dots,
\end{align}
where $\sigma_0^{t\bar{t} }$ is the Born-level production cross section, and $\mathcal{H}_i \equiv \langle \bm{\mathcal{H}}_i \rangle$.
In the quark annihilation channel, the two dipoles need to be showered separately.

The remainder of this appendix is organized as follows. The set up of the shower is discussed in Section~\ref{sec:app_setUpShower} by looking at an explicit example. The shower procedure is discussed in \ref{sec:app_startShower}, also in this case an explicit event is considered as an example. 
Finally, a brief outline of the algorithm is provided in Section~\ref{sec:app_alg}.

\subsection{Interface to LHE files}\label{sec:app_setUpShower}
The LHEF produced by {\sc MadGraph5\Q{_}aMC@NLO} contains, for each event,  a list of particles with their momenta and information about the nature of the particles. One of these events is listed below. For simplicity, only  the particle id, status, colors and four-momenta are provided here:

\begin{center}
\begin{tabular}{rrllllll}
{\tt id } & {\tt s} & {\tt c1} & {\tt c2} & {\tt x mom } & {\tt y mom} & {\tt z mom} & {\tt energy}  \\
\hline
{\tt 21} & {\tt -1} & {\tt 503} & {\tt 502} & {\tt +0.000e+00} & {\tt +0.000e+00} & {\tt +9.106e+01} & {\tt 9.106e+01}  \\
{\tt 21} & {\tt -1} & {\tt 501} & {\tt 503} & {\tt -0.000e+00} & {\tt -0.000e+00} & {\tt -6.834e+02} & {\tt 6.834e+02}  \\
{\tt 6} & {\tt 2} & {\tt 501} & {\tt 0} & {\tt +1.256e+02} & {\tt+8.244e+01 } & {\tt -4.504e+02} & {\tt 5.047e+02}  \\
{\tt 24} & {\tt 2} & {\tt 0} & {\tt 0} & {\tt +1.165e+02} & {\tt -4.050e+00} & {\tt -2.698e+02} & {\tt 3.042e+02}  \\
{\tt -6} & {\tt 2} & {\tt 0} & {\tt 502} & {\tt -1.256e+02} & {\tt -8.244e+01} & {\tt -1.419e+02} & {\tt2.697e+02}  \\
{\tt -24} & {\tt 2} & {\tt 0} & {\tt 0} & {\tt -2.846e+01} & {\tt -6.129e+00} & {\tt -5.262e+00} & {\tt 8.540e+01}  \\
{\tt 14} & {\tt 1} & {\tt 0} & {\tt 0} & {\tt +4.077e+01} & {\tt +2.321e+01} & {\tt -4.318e+01} & {\tt 6.377e+01}  \\
{\tt -13} & {\tt 1} & {\tt 0} & {\tt 0} & {\tt +7.575e+01} & {\tt -2.726e+01} & {\tt -2.266e+02} & {\tt 2.404e+02}  \\
{\tt -12} & {\tt 1} & {\tt 0} & {\tt 0} & {\tt +2.455e+01} & {\tt -1.372e+01} & {\tt +9.769e+00} & {\tt 2.978e+01}  \\
{\tt 11} & {\tt 1} & {\tt 0} & {\tt 0} & {\tt -5.301e+01} & {\tt +7.597e+00} & {\tt -1.503e+01} & {\tt 5.562e+01}  \\
{\tt 5} & {\tt 1} & {\tt 501} & {\tt 0} & {\tt +9.077e+00} & {\tt +8.649e+01} & {\tt -1.806e+02} & {\tt 2.005e+02}  \\
{\tt -5} & {\tt 1} & {\tt 0} & {\tt 502} & {\tt -9.715e+01} & {\tt -7.631e+01} & {\tt -1.366e+02} & {\tt 1.842e+02}  
\end{tabular}
\end{center}
 
This particular event consists of two incoming (\verb|status: -1|) gluons (\verb|id: 21|), four intermediate particles (\verb|status: 2|), namely a top quark (\verb|id: 6|), a $W^+$-boson (\verb|id: 24|) and their antiparticles (denoted with a negative \verb|id|), and the six final-state particles (\verb|status: 1|) $\nu_\mu$ (\verb|id: 14|), $\mu^+$ (\verb|id: -13|), $\bar{\nu}_e$ (\verb|id: -12|), $e^-$ (\verb|id: 11|) and the $b\bar{b}$-pair (\verb|id: (-)5|).
The color-connection indices $c_1$ and $c_2$ will be explained below. For illustrative purposes in the rest of this appendix, calculations are carried out by rounding to three digits after the decimal point.

The final state leptons in the event must satisfy the cuts listed in Table 1 of \cite{ATLAS:2012al}. The momenta of the $b\bar{b}$ pair are needed, because they define the direction of the $b$-jets, which are cut out of the gap region (or veto region), see \eqref{eq:outReg} and Figure \ref{fig:outsideRegion}. 

\begin{figure}[t!]
	\centering
	\begin{overpic}[width=0.35\textwidth]{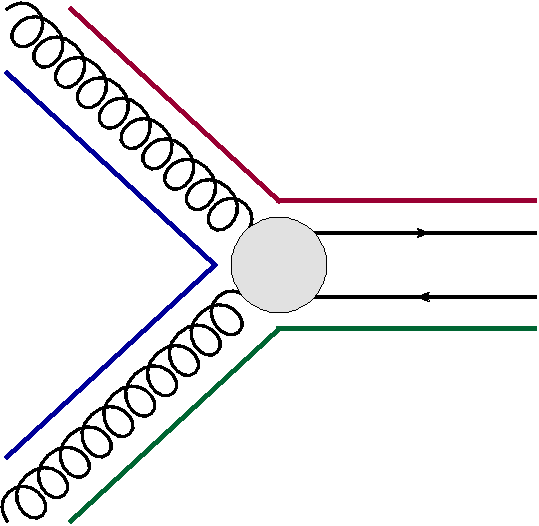}
		\put(102,51){$ t$}
		\put(102,41){$ \bar{t}$}
		\put(-7,100){$ g_2$}
		\put(-7,-5){$ g_1$}
		\put(102,59){\color[RGB]{153,0,31}501}
		\put(9,99){\color[RGB]{153,0,31}501}
		\put(-13,83){\color[RGB]{0,0,153}503}
		\put(-13,8){\color[RGB]{0,0,153}503}
		\put(103,33){\color[RGB]{0,102,51}502}
		\put(9,-6){\color[RGB]{0,102,51}502}
	\end{overpic}
	\caption{Color dipoles in the sample top-pair production event discussed in text. \label{fig:dipolestructure_App}}
\end{figure}
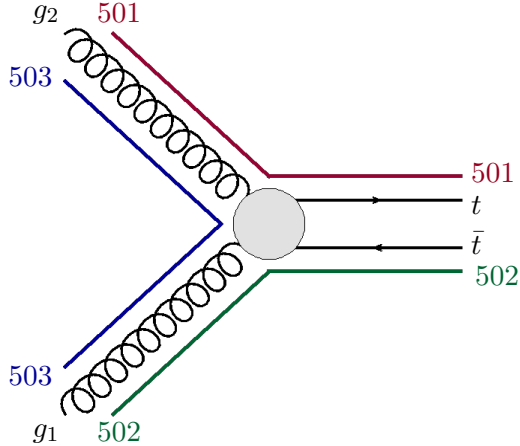

In the following we  illustrate the shower algorithm with the $t\bar{t}$ production process. The additional dipoles arising from the top-quark decay can be showered exactly in the same way.  In the end, it is necessary to multiply the results of each shower to get the complete result of each event, see \eqref{eq:app_multOfDip}.
The showering process of the dipoles associated to the top-pair production starts by selecting the momenta of the initial-state partons and  the momenta of the $t\bar{t}$ pair. These momenta are stored in dipoles according to the large-$N_c$ color information assigned by {\sc MadGraph5\Q{_}aMC@NLO}. In the sample event depicted in Figure \ref{fig:dipolestructure_App}, the color index associated to the top quark is {\tt c1: 501}, while the color indices associated to the gluon in the second line of the list above ($g_2$) are {\tt c1: 501, c2: 503}. The color indices of the gluon in the first line of the list ($g_1$) are {\tt c1: 503, c2: 502} and the color index of the anti-quark is {\tt c2: 502}. The color indices indicate which lines are color connected and the shower algorithm orders the particles in such a way that equal color indices are adjacent to each other; so that the list of the color indices is {\tt 501,501,503,503,502,502}. Therefore the ordering of the particles is ($t$,$g_2$,$g_1$,$\bar{t}$) which represents a dipole with two intermediate gluons.

Since the algorithm only requires information about the direction of the particles, we normalize the components of the momenta to their energy. These normalized momenta are stored in an array
\begin{align}
\{ \underline{u} \} =\left\{ \frac{p_t}{E_t},\frac{p_{g_2}}{E_{g_2}},\frac{p_{g_1}}{E_{g_1}},\frac{p_{\bar{t}}}{E_{\bar{t}}}\right\}=\left\{\left(\begin{matrix} 1\\0.249\\ 0.163\\-0.892\end{matrix}\right),\left(\begin{matrix} 1\\0\\ 0\\-1\end{matrix}\right),\left(\begin{matrix} 1\\0\\ 0\\1\end{matrix}\right),\left(\begin{matrix} 1\\-0.466\\ -0.306\\ -0.526\end{matrix}\right)\right\}\,  \label{eq:eventlist}
\end{align}
such that each adjacent pair of vectors represents a color dipole.

We then calculate the virtual correction of each dipole using \eqref{eq:virtCorr} by boosting the 
two vectors in each dipole into a frame where they are back-to-back, as explained in  Section \ref{sec:evaluation},
and by subsequently evaluating the velocities $\beta_i^\prime,\beta_j^\prime$. Each dipole contributes to the virtual corrections a factor 
\begin{align} \label{eq:app_Vij}
V_{ij}= 4 N_c \left(
\frac{1 + \beta^\prime_i \beta^\prime_{j}}{\beta^\prime_i + \beta^\prime_{j} }\left(y_{\rm max}-y_{\rm min}\right) -\frac{1}{2}\left(\delta_{v_i}+\delta_{v_{j}}\right)\right) ,
\end{align}
where  $N_c=3$.

Let us illustrate the process by explicitly calculating  the virtual corrections for the first dipole in the list, $V_{12}$, according to the method put forward in Section \ref{sec:evaluation}. The first step consists in boosting the two normalized momenta in the rest frame.
The sum of the two normalized momenta is 
\begin{align}
U=u_1+u_2=\left(\begin{matrix} 2\\0.248\\ 0.163\\-1.892\end{matrix}\right).
\end{align}
This vector is then aligned to the $x$-axis by means of the matrix 
\begin{align}
X =  \left(
\begin{array}{cccc}
\; 1&\; 0 &\;0 &\; 0\\
\; 0 &\; 0.130 &\; 0.085  &\;-0.988  \\
\; 0 &\; 0.085  &\; 0.992 &\;  0.097 \\
\; 0&\; -0.988&\; 0.097 &\; -0.122
\end{array}
\right),
\end{align}
which leads to the vectors 
\begin{align} \label{eq:checkU}
\check{ U }\equiv\left(\begin{matrix} 2\\1.916\\ 0\\0\end{matrix}\right),\hspace{1cm}\check{ u }_1=\left(\begin{matrix} 1\\0.928\\ 0.097\\-0.122\end{matrix}\right),\hspace{1cm}\check{ u }_2=\left(\begin{matrix} 1\\0.987\\ -0.097\\0.122\end{matrix}\right),
\end{align}
with $\check{ U }$ having vanishing $y$ and $z$ components by construction.

For the dipole under consideration the factor $\beta$ introduced in (\ref{eq:totMom}) is $\beta = 0.958$, which leads to boost matrix $B$ in (\ref{eq:matB}) with the following entries
\begin{align}
B =  \left(
\begin{array}{cccc}
\; 3.481&\; -3.334 &\;0 &\; 0\\
\; -3.334 &\;  3.481 &\; 0  &\; 0  \\
\; 0 &\; 0  &\; 1 &\; 0 \\
\; 0&\; 0&\; 0 &\; 1
\end{array}
\right).
\end{align}
By applying the matrix $B$ to the vectors in (\ref{eq:checkU})
one finds
\begin{align} \label{eq:tildeU}
\tilde{ U }=\left(\begin{matrix} 0.575\\0\\ 0\\0\end{matrix}\right),\hspace{1cm}\tilde{ u }_1=\left(\begin{matrix} 0.387\\-0.104\\ 0.097\\-0.122\end{matrix}\right),\hspace{1cm}\tilde{ u }_2=\left(\begin{matrix} 0.187\\0.104\\ -0.097\\0.122\end{matrix}\right) ,
\end{align}
with $\vec{\tilde{u}}_1=-\vec{\tilde{u}}_2$ and $\vec{\tilde{ U}}=0$ by construction.

Finally one needs to apply the rotation matrix that aligns the two vectors along the $z$ axis (\ref{eq:matZ})
\begin{align}
Z =  \left(
\begin{array}{cccc}
\; 1&\; 0 &\;0 &\; 0\\
\; 0 &\; 0.811 &\; 0.175  &\; -0.558  \\
\; 0 &\; 0.175  &\; 0.838 &\; 0.517 \\
\; 0&\; -0.558&\; 0.517 &\; -0.649
\end{array}
\right)  .
\end{align}
By applying the matrix $Z$ to the vectors in (\ref{eq:tildeU}) one finds
\begin{align}
U^\prime=\tilde{ U }\,,\hspace{1cm}u^\prime_1=\left(\begin{matrix} 0.387\\0\\ 0\\0.187\end{matrix}\right)\,,\hspace{1cm}u^\prime_2=\left(\begin{matrix} 0.187\\0\\ 0\\-0.187\end{matrix}\right).
\end{align}
In this frame we obtain the factors $\beta'_1,\, \beta'_2,\, y_{{\rm min}},\, y_{{\rm max}} $ that are needed in the calculation of the virtual corrections
\begin{equation}
\beta^\prime_1=0.483 \, , \hspace{2cm}\beta^\prime_2=1\,, \nonumber
\end{equation}
\begin{equation} 
y_{{\rm min}}=-y_{{\rm cut}}=-4.184 \, ,\hspace{2cm}y_{{\rm max}}= \frac{1}{2}\ln \left(\frac{1+\beta^\prime_2}{1-\beta^\prime_1}\right)=0.677\, ,\label{eq:app_yminymax}
\end{equation}
and the full boost and inverse boost 
\begin{align}
L = \left(
\begin{array}{cccc}
\; 3.481&\; -0.433 &\; -0.284 &\; 3.293\\
\; -2.705 &\; 0.933 &\; 0.360  &\; -2.704  \\
\; -0.583 &\; -0.360  &\; 0.933 &\; -0.583 \\
\; 1.860&\;  0.433 &\; 0.284 &\; 2.047
\end{array}
\right),\hspace{2cm}
L^{-1} = g \, L^T \, g\,, \label{eq:app_boosts}
\end{align}
with $g={\rm diag}(1,-1,-1,-1)$.
Let us briefly explain the value of $y_{{\rm cut}}$ which cuts off the collinear divergence arising for massless partons. Following (A.1) of \cite{Balsiger:2018ezi}, we obtain the value of the cut-off by imposing a rapidity cut  $\eta_{\rm cut}$ in the lab frame and then computing the value of $y_{{\rm cut}}$ that this corresponds to in the center-of-mass frame. This leads to the somewhat complicated expression
\begin{equation}\label{eq:cutoff}
y_{{\rm cut}} = \ln\! \left(\cos\Big(\frac{\theta }{2}\Big) + \sqrt{\cos^2\Big(\frac{\theta }{2}\Big) + \sin^2\Big(\frac{\theta}{2}\Big) e^{2 \eta_{{\rm cut}}}}\right),
\end{equation}
where $\theta$ is the angle between the two vectors forming the dipole in the lab frame. One immediately sees that $y_{{\rm cut}}=\eta_{{\rm cut}}$ for back-to-back vectors $\theta=\pi$.
The value $y_{{\rm cut}}$ for the dipole under consideration is obtained after setting $\eta_{{\rm cut}}=6$.

Since  the top quark is massive and the gluon is massless, one has $\delta_{v_1}=1$ and $\delta_{v_2}=0$ in calculating the contribution of this dipole to the virtual corrections (\ref{eq:app_Vij}). The virtual correction associated to the first dipole in the event is given by $V_{12}=52.332$. The value of the virtual corrections in (\ref{eq:app_Vij}) for each of the three dipoles in the event is stored in another array 
\begin{equation}
\{\underline{V} \}=\left(52.332,144.000,78.443\right).
\end{equation}

\subsection{Monte Carlo implementation of $\mathcal{H}_i(t)$}\label{sec:app_startShower}

As outlined in the introduction of this appendix, we could also have set up the showering of each of these three dipoles individually and then multiplied the results. To reduce computation time, we treat color-connected dipole structures such as \eqref{eq:eventlist} as a single dipole which has already emitted two gluons at $t=0$. It is convenient to multiply \eqref{eq:Rttb} by the virtual correction and to define
\begin{align}
\hat{R}_{t\bar{t}}(t)= V_{4}\, R_{t\bar{t}}(t)=\mathcal{\hat{H}}_4(t) + \int \frac{d\Omega_5}{4\pi} \mathcal{\hat{H}}_{5}(t) +\int \frac{d\Omega_5}{4\pi}\int \frac{d\Omega_6}{4\pi}  \mathcal{\hat{H}}_{6}(t) + \dots\,,\label{eq:app_sigmaLL}
\end{align}
where the hat indicates a multiplication of the hard functions by the total virtual correction of the four legs
\begin{equation} \label{eq:app_Vtot}
V_{4} \equiv V_{\rm tot} = V_{12} + V_{23} + V_{34} \, , 
\end{equation}
and division by the LO cross section, see \eqref{eq:Rttb}. The term $\mathcal{H}_4(t)$ corresponds to the initial tree-level configuration with associated dipole-structure \eqref{eq:eventlist}, while $\mathcal{H}_5(t)$ contains an additional gluon emitted from one of the dipoles in \eqref{eq:eventlist}.

Let us now analyze the individual terms in (\ref{eq:app_sigmaLL}). The first term $\mathcal{\hat{H}}_4(t)$ denotes the evolution from the hard scale to the low scale without any radiation from any of the dipoles (compare to the first line of \eqref{eq:iterRG}) and is given by
\begin{align}
\mathcal{\hat{H}}_4(t)&= V_{4} \frac{\mathcal{H}_4(0)}{\sigma_0^{t \bar{t}}} e^{-t V_{4} } = V_{4} e^{-t  V_{4}}\, ,
\end{align} 
since by definition $\mathcal{\hat{H}}_4(0)=R_{t\bar{t}}(0)=1$. At this stage it is convenient to define the probability distribution
\begin{align}
\mathcal{P}(V,t)=V\,e^{-V\,t}
\end{align} 
such that $\mathcal{\hat{H}}_4(t)=\mathcal{P}(V_{4},t)$.

The function $\hat{\mathcal{H}}_5(t)$ consists of the initial four hard partons plus one additional parton emitted by any of the dipoles at an evolution time earlier than $t$. In the large-$N_c$ limit, the new emission occurs from any one of the three dipoles in the list so that we get three terms
\begin{align}\label{eq:H5HatTAll}
\mathcal{\hat{H}}_5(t)&= \mathcal{\hat{H}}^{(1)}_5(t) + \mathcal{\hat{H}}^{(2)}_5(t) + \mathcal{\hat{H}}^{(3)}_5(t)\nonumber \\
& =  \int_{0}^{t} dt' \,\mathcal{\hat{H}}_{4}(t')  \Big(
R_{12}^5 e^{-(t-t')   V_5^{(1)}}
+R_{23}^5 e^{-(t-t') V_5^{(2)}} +R_{34}^5 e^{-(t-t') V_5^{(3)}}\Big) \, ,
\end{align}
each evolving with its specific virtual correction
\begin{align}
V_5^{(1)} &= V_{15}+V_{52}+V_{23}+V_{34}\, , \nonumber \\
V_5^{(2)} &= V_{12}+V_{25}+V_{53}+V_{34}\,  , \nonumber \\
V_5^{(3)} &= V_{12}+V_{23}+V_{35}+V_{54} \,. 
\end{align}
The quantity $R_{ij}^5$ corresponds to the real correction factor as given in {\eqref{eq:realCorr}} when the fifth parton  is emitted in the direction $n_5$ from the dipole of legs $i$ and $j$
\begin{align}
R_{ij}^5 & = 4 N_c  \widetilde{W}_{ij}^5 \Theta_{\rm in}(n_{5}) \,.\label{eq:app_realEmission}
\end{align}
To bring \eqref{eq:H5HatTAll} into a form suitable for Monte Carlo implementation, we now strategically insert factors of one. We rewrite the first term on the right-hand side of the first line in the form
\begin{align}
\mathcal{\hat{H}}^{(1)}_5(t) &=\int_{0}^{t} dt' \,\mathcal{\hat{H}}_{4}(t') \frac{R_{12}^5}{V_{12}}\frac{V_{12}}{V_4}\frac{V_{4}}{V_5^{(1)} } V_5^{(1)}  e^{-(t-t')  V_5^{(1)} } \nonumber \\
&=\int_{0}^{t} dt' \,\mathcal{P}( V_4,t')  \frac{R_{12}^5}{V_{12}}\frac{V_{12}}{V_4}\frac{V_4}{ V_5^{(1)}} \mathcal{P}( V_5^{(1)} ,t-t') \, . \label{eq:H5HatT}
\end{align} 
The integral $\int d\Omega_5/4\pi$, over the direction of the emission in \eqref{eq:app_sigmaLL}, is evaluated by Monte Carlo methods.  The factor $V_{12}/V_4$ is the weight of the dipole $(12)$ in the total virtual correction $V_4$ and is interpreted as the probability of having an emission from the dipole $(12)$. 

For the case of $\mathcal{\hat{H}}^{(1)}_5(t)$ the sets of direction vectors and of associated virtual corrections are
\begin{align}
\{ \underline{u}\}&=\left(u_t,n_5,u_{g_2},u_{g_1},u_{\bar{t}}\right) , & \{\underline{V} \} &=\left(V_{15},V_{52},V_{23},V_{34}\right)  .
\end{align}
The contribution of terms arising from $\mathcal{\hat{H}}^{(1)}_5(t)$ to $\mathcal{\hat{H}}_6(t)$ in \eqref{eq:app_sigmaLL} is denoted by
\begin{align}
\mathcal{\hat{H}}^{(1)}_6(t)&= \mathcal{\hat{H}}^{(11)}_6(t) + \mathcal{\hat{H}}^{(12)}_6(t)  + \mathcal{\hat{H}}^{(13)}_6(t) +  \mathcal{\hat{H}}^{(14)}_6(t) \, .
\end{align}
It involves four terms $\mathcal{\hat{H}}^{(1i)}_6(t) $, where $i$ denotes the position of the dipole which makes the next emission. These terms have the same structure as the ones in the first emission. For example
\begin{align}\label{eq:H6HatT}
 \mathcal{\hat{H}}^{(12)}_6(t) &= \int_{0}^{t} dt'' \,\mathcal{\hat{H}}^{(1)}_{5}(t'') R_{52}^6\, e^{-(t-t'') V_6^{(12)}} \, ,
 \end{align}
 where the relevant virtual contribution is given by
 \begin{equation}
V_6^{(12)} = V_{15}+ V_{56}+ V_{62}+ V_{23} +V_{34} \, .
\end{equation}
One can rewrite the quantity in \eqref{eq:H6HatT} by strategically inserting factors of one, as it was done in  \eqref{eq:H5HatT}, to find
\begin{align} 
 \mathcal{\hat{H}}^{(12)}_6(t) =\int_{0}^{t} dt'' \,\mathcal{\hat{H}}^{(1)}_{5}(t'') \frac{R_{52}^6}{V_{52}} \frac{V_{52}}{V_5^{(1)}}\frac{V_5^{(1)}}{V_6^{(12)}} \mathcal{P}( V_6^{(12)}, t-t'' ) \, .
\end{align}
This iterative procedure  can be repeated to calculate all of the terms $\mathcal{\hat{H}}_{i}(t)$.

In the parton shower algorithm, this procedure is implemented  as follows. For the numerical example, we again consider the event already set up for the showering in Appendix~\ref{sec:app_setUpShower}. At first, the shower is initialized as follows:
\begin{equation}
t=0 \, , \qquad  V_{{\rm tot}}= V_4 = V_{12} + V_{23} + V_{34} = 274.765 \, , \qquad w=\frac{1}{n_{sh}}\,.
\end{equation}
The initial weight $w$ of an individual event is the inverse of the number of showerings of a tree-level event in the LHE file, $n_{sh}$; additional weight factors arising from integrands such as the one in \eqref{eq:H5HatT} are discussed below.

At first it is necessary to randomly generate a time step $\Delta t$ according to the  probability density $\mathcal{ P} (V_{{\rm tot}},\Delta t)$. We generate random time steps $\Delta t$ according to this distribution by taking the cumulant $u \equiv \mathcal{P}/V_{{\rm tot}}\in[0,1]$, inverting it
\begin{align}
\Delta t=-\frac{\ln u}{V_{{\rm tot}}}\, ,
\end{align}
and using an equally distributed random variable $u\in [0,1]$.
For our sample event, we choose $u=0.5$ for illustration, which yields $\Delta t = 0.00252$. To account for $\mathcal{\hat{H}}_4(\Delta t) = \mathcal{ P} (V_4,\Delta t)$, we add a weight $w$ into a histogram in a bin corresponding to the  time $t^{{\rm new}} = \Delta t= 0.00252$. Once the shower will be finished, this histogram will provide the gap fraction $\hat{R}_{t\bar{t}}(t)$. 

Next, we use \eqref{eq:H5HatTAll} to iteratively compute $\mathcal{\hat{H}}_5(t)$ at a time $t>t^{{\rm new}}$. To do so, we interpret $t^{{\rm new}}$ as the time at which one of the three original dipoles emits a parton. Looking at \eqref{eq:H5HatTAll}, one sees that $\mathcal{\hat{H}}_5(t)$ has three terms. 
The first of these terms is given in \eqref{eq:H5HatT}; it involves a product of several factors. The first factor is the probability density $\mathcal{P}(V_4,t')$ which was taken into account when generating the time step $\Delta t$. Showering multiple times, one gets a Monte Carlo approximation of the integral over $dt'$. We then have the three factors ${R_{12}^5}/{V_{12}}$, ${V_{12}}/{V_4}$ and $V_4/ V_5^{(1)}$.  The last two of these factors  can be treated as probabilities since the value of these ratios is always in the interval $[0,1]$. The factor ${R_{12}^5}/{V_{12}}$ corresponds to the phase-space integral and will be treated as a weight. The very last factor $\mathcal{P}( V_5^{(1)} ,t-t')$ in \eqref{eq:H5HatT}  represents the emission probability in the time interval $t-t'$.  

The factor $V_{12}/V_4$ in \eqref{eq:H5HatT} is the weight of the dipole $(12)$ in the total virtual correction $V_4$ and is interpreted as a probability of selecting the dipole $(12)$ for the emission.The shower algorithm selects one of the terms $\mathcal{\hat{H}}^{(i)}_5(t)$ according to the probabilities $\{\underline{V}\}/V_{\rm tot} = \{ V_{12}, V_{23},V_{34}\} / V_4$. To implement this, one can draw a random value $u\in[0,1]$. Then, the $i$-th dipole is assumed to emit, if the cumulative sum of the virtual corrections of all the dipoles from $1$ through $i$ divided by the total virtual corrections is smaller than $u$. In the example under consideration, this means that if $u<0.190=V_{12}/V_{{\rm tot}}$ it is the first dipole that emits, if $0.190<u<0.715=(V_{12}+V_{23})/V_{{\rm tot}}$ the emission arises from the second dipole, and if $0.715<u$ the third dipole emits. For the purposes of this discussion, let us assume  that $u=0.1$, such that first dipole emits.

Now that the algorithm determined which dipole radiates, one can boost into  the back-to-back frame of  the selected dipole. The procedure to do this for the first dipole was already illustrated in Appendix~\ref{sec:app_setUpShower}.
In this frame, the algorithm draws two more random numbers, namely 
\begin{align}
\phi'&= 2\pi u_\phi\, , \nonumber \\
y'&=y_{{\rm min}}+u_y\;(y_{{\rm max}}-y_{{\rm min}}) \, , \label{eq:yprime}
\end{align}
with $u_i\in[0,1]$ and the integration boundaries $y_{{\rm min}}$ and $y_{{\rm max}}$ as given previously in \eqref{eq:app_yminymax}. For illustration, assume that $u_\phi=u_y=0.5$, which yields $\phi^\prime=\pi$ and $y^\prime=-1.754$. With this input one then obtains the four-vector of the newly emitted parton as
\begin{align}
n^\prime_5=\left(\begin{matrix} 1\\n_T \cos \phi'\\ n_T \sin \phi'\\ n_z \end{matrix}\right) = \left(\begin{matrix} 1\\-0.293\\0\\ -0.956\end{matrix}\right) , \label{eq:n5pp}
\end{align}
with $n_z =(e^{2y'} -1 )/(\beta_1^\prime e^{2y'} + \beta_2^\prime)$ and $n_T = \sqrt{1-n_z^2}$. 

The new vector is then boosted back to the lab frame by using the inverse boost matrix $L^{-1}$ given in \eqref{eq:app_boosts}. Subsequently, the vector is normalized in such a way that the energy component is  $1$:
\begin{align}
p_5=L^{-1}n_5^\prime=\left(\begin{matrix} 4.466\\-0.254\\-0.093\\ -4.458\end{matrix}\right)\Rightarrow n_5=\left(\begin{matrix} 1\\-0.057\\-0.021\\ -0.998\end{matrix}\right)  .
\end{align} 
With the new vector $n_5$, the algorithm evaluates the factor 
\begin{align}
\frac{R_{12}^5}{V_{12}}=  \frac{4 N_c \frac{1+\beta_1^\prime \beta_2^\prime}{\beta_1^\prime+\beta_2^\prime}\left(y_{{\rm max}}-y_{{\rm min}}\right)\left(1-\frac{1}{2}\frac{W_{11}^5}{W_{12}^5}\right)}{V_{12}}=1.106\, . \label{eq:app_R12V12}
\end{align}
The quantity in \eqref{eq:app_R12V12} is strictly positive (as shown in Section \ref{sec:posDef}), but it can not be treated as a probability, since it can exceed unity.\footnote{This arises because we include the monopoles as a weight factor into the dipole integral. Alternatively, one could integrate the full, modified dipole $\widetilde{W}_{ij}^k$ given in \eqref{eq:realinteg}. The integral can be done and leads to a more complicated version of the rapidity variable \eqref{eq:genrap}. However, the inversion to the angle $\theta$ can then not be done analytically, in contrast to \eqref{eq:genrap}.} Therefore the algorithm accounts for it by modifying the weight factor
\begin{align}
w^{{\rm new}}=w\, \frac{R_{12}^5}{V_{12}} \, .
\end{align}
We have to ensure that the real emission is not in the veto region, as imposed by the $\Theta_{\rm in}(n_{5})$-function in \eqref{eq:app_realEmission}. This is done by checking the conditions \eqref{eq:outReg} with $q=n_5$. We obtain
\begin{align}
\Delta R(b,n_5)=2.860 \,, \hspace{2cm} \Delta R(\bar{b},n_5)=2.561 \,, \hspace{2cm} y(n_5)=-3.496 \, ,
\end{align}
implying that the new emission fulfills all the conditions $\Theta_{\rm in}(n_{5})$ and one can proceed to the next step of the algorithm. If the condition would have not been fulfilled, the shower would have been stopped at this point and the algorithm would have restarted at the beginning, after erasing all information on this showering other than the histogram  entry for $\mathcal{\hat{H}}_4(t)$ at $t=\Delta t$.

Since the generated vector $n_5$ was not in the gap region, the algorithm continues by adding the new vector to the list of vectors in between the first and second vector of the list in \eqref{eq:eventlist}.
In addition, the algorithm updates the virtual correction list by replacing the virtual correction $V_{12}$ by $V_{15}$ and $V_{52}$:
\begin{equation}
\{\underline{u}\}=\left\{\left(\begin{matrix} 1\\0.248\\ 0.163\\-0.892\end{matrix}\right),\left(\begin{matrix} 1\\-0.057\\-0.021\\ -0.998\end{matrix}\right), \left(\begin{matrix} 1\\0\\ 0\\-1\end{matrix}\right),\left(\begin{matrix} 1\\0\\ 0\\1\end{matrix}\right),\left(\begin{matrix} 1\\-0.466\\ -0.306\\ -0.526\end{matrix}\right)\right\} ,
\end{equation}
\begin{equation}
\{\underline{V}\}=\left\{54.806,61.974,144.000,78.433\right\} \Rightarrow V_{{\rm tot}}^{\rm new}= V_5^{(1)}=339.212\, .
\end{equation}

There is one last factor in $\mathcal{\hat{H}}^{(1)}_5(t)$ that was not accounted for so far, namely the factor $V_{4}/V_5^{(1)}$.  This last factor can again be treated as a probability, as the new virtual corrections are always larger than the old ones.\footnote{Whether this is true depends on the form of the angular cutoff used in the shower, see \cite{Balsiger:2018ezi}.}  This means that instead of multiplying the weight by this factor, one can again draw a random variable $u\in(0,1)$ and if $u<0.810=V_{4}/V_5^{(1)}= V_{{\rm tot}}^{{\rm old}}/V_{{\rm tot}}^{{\rm new}} $, the algorithms continues with the generation of a time step starting at $t=t^{{\rm new}}$ with weight $w=w^{{\rm new}}$. In the opposite case, the shower is stopped. One could also treat $V_{4}/V_5^{(1)}$ as a weight factor, but, due to the iterative nature of the shower, the weights of the individual steps multiply each other, leading to events with small weight which would render the shower inefficient.

The new time step is generated in exactly in the same way as before according to  the new probability density $\mathcal{ P} (V_{{\rm tot}}^{{\rm new}},\Delta t')$ and completes the calculation of $\mathcal{\hat{H}}_5(t)$ by  writing the weight into the histogram at $t=\Delta t +\Delta t'$. After this is done, the algorithm proceeds  to calculate $\mathcal{\hat{H}}_6(t)$. Looking at \eqref{eq:H6HatT}, one sees that the same procedure described for the calculation of $\mathcal{\hat{H}}_5(t)$ can be used, since it involves the same type of ingredients:
\begin{itemize}
 \item[a)] An emitting dipole is chosen, each with probability of ${V_{ij}}/{V_{{\rm tot}}}$.
 \item[b)]  The emission is generated and the factor ${R_{ij}^6}/{V_{ij}}$ is calculated and multiplied to the weight. 
 \item[c)] If the emission is not in the veto region, the algorithm proceeds with probability ${V_{{\rm tot}}}/{V_{{\rm tot}}^{{\rm new}}}$ and and generates a new time step using $\mathcal{ P} ( V_{{\rm tot}}^{{\rm new}},\Delta t'')$, which gives $\mathcal{\hat{H}}_6(t)$ with  $t = \Delta t+\Delta t'+\Delta t''$.
\end{itemize}
The iterative calculation of all the  $\mathcal{\hat{H}}_{i}(t)$ with $i > 6$ can be carried out in the same way until one reaches the necessary maximal value for $t$ determined by the lowest value of $\mu_s\sim Q_0$ in the problem under consideration. Each showering generates several hard functions at successively larger times until it terminates. In the calculations presented in this work, we used an upper limit of $t_{\rm max}=0.1$, which corresponds to $\mu_s\approx 0.75\,{\rm GeV}$ after applying the profile \eqref{eq:profFun} and $\mu_h=150\,{\rm GeV}$. 

\subsection{Parton shower algorithm}\label{sec:app_alg}

This section summarizes the different steps in the shower algorithm, which were discussed in detail in Appendix~\ref{sec:app_startShower} in the context of the showering of a particular event.
The shower algorithm described below is applied $n_{sh}$ times to each tree-level event. In the following, we describe one such shower event. For the results presented in our paper, we used about $10^5$ tree-level events and worked with $n_{sh}=10^4$.

\subsubsection*{Step 0. Set up the shower}
Store all Wilson-line directions according to their color information into an array, and calculate the virtual corrections of each dipole
\begin{align}
\{\underline{u}\}&=\left\{u_1,u_2,\dots,u_m\right\} ,\nonumber\\
\{\underline{V}\}&=\left\{V_{12},V_{23},\dots,V_{(m-1)m}\right\} ,\label{eq:initialVrecipe}
\end{align} 
where
\begin{equation}
V_{ij}= 4 N_c \left(
\frac{1 + \beta^\prime_i \beta^\prime_{j}}{\beta^\prime_i + \beta^\prime_{j} }\left(y_{\rm max}-y_{\rm min}\right) -\frac{1}{2}\left(\delta_{v_i}+\delta_{v_{j}}\right)\right).
\end{equation}
An expression for the rapidity values can be found in \eqref{boundary}. Initiate the shower algorithm with the initial settings
\begin{equation}
t=0 \, ,\qquad V_{{\rm tot}}=\sum_i V_i \, , \qquad w=\frac{1}{n_{sh}} \, . \label{eq:initialValuesRecipe}
\end{equation}

\subsubsection*{Step 1. Generate time step}
Generate a random number $u\in[0,1]$ and calculate 
\begin{equation}
\Delta t = -\frac{\ln(u)}{ V_{{\rm tot}}} \, ,
\end{equation}
which is added to the variable of the evolution time
\begin{equation}
t \rightarrow t=t+\Delta t \, .
\end{equation}

\subsubsection*{Step 2. Insert weight into histogram}
At this new time $t$, insert $w$ into the histogram.

\subsubsection*{Step 3. Choose emitting dipole}
Randomly choose a dipole which emits the next emission, where each dipole with legs $i$ and $j$ emits with probability 
\begin{align}
p_{ij}=\frac{V_{ij}}{V_{{\rm tot}}} \, .
\end{align}

\subsubsection*{Step 4. Create emission}
Boost to the frame where the emitting legs are back-to-back along the $z$-axis and in that frame choose an emission direction $n_k^\prime$ by generating a random angle $\phi' \in [0,2\pi]$  and rapidity $y' \in [y_{{\rm min}},y_{{\rm max}}]$. Boost $n_k^\prime$ back into the lab frame and normalize it as $n_k=(1,\vec{n}_k)$. Update the event weight according to
\begin{equation}
w\rightarrow w=w \frac{R_{ij}^k}{V_{ij}}=w \frac{4 N_c \frac{1+\beta_i^\prime \beta_j^\prime}{\beta_i^\prime+\beta_j^\prime}\left(y_{{\rm max}}-y_{{\rm min}}\right)\left(1-\frac{1}{2}\frac{W_{ii}^k+W_{jj}^k}{W_{ij}^k}\right)}{V_{ij}}\, .
\end{equation}

\subsubsection*{Step 5a. Emission not in veto region}
If the emission $n_k$ is in the allowed region, add the new direction to the list $\{\underline{u}\}$ of particles in the event and replace the virtual corrections between legs $i$ and $j$ by the two virtual corrections between the legs $i$, $k$ and $k$, $j$:
\begin{align}
\{\underline{u}\}=\left\{u_1,\dots,u_i,u_j,\dots,u_m\right\}&\Rightarrow \{\underline{u}\}=\left\{u_1,\dots,u_i,n_k,u_j,\dots,u_m\right\}  , \nonumber \\
\{\underline{V}\}=\left\{V_{12},\dots,V_{ij},\dots,V_{(m-1)m}\right\}&\Rightarrow \{\underline{V}\}=\left\{V_{12},\dots,V_{ik},V_{kj},\dots,V_{(m-1)m}\right\}.
\end{align}
Go back to step 1 with a probability
\begin{align}
\frac{V_{{\rm tot}}}{V_{{\rm tot}}^{{\rm new}}}:=\frac{V_{12}+\dots+V_{ij}+\dots+V_{(m-1)m}}{V_{12}+\dots+V_{ik}+V_{kj}+\dots+V_{(m-1)m}}\, .
\end{align}
The algorithm restarts from step 1 with the new arrays $\{\underline{u}\}$, $\{\underline{V}\}$ and $V_{{\rm tot}}=V_{{\rm tot}}^{{\rm new}}$.

With the probability of $1-({V_{{\rm tot}}}/{V_{{\rm tot}}^{{\rm new}}})$ the shower is stopped. Of course one can also set an upper limit $t_{\rm max}$ on $t$ after which the shower stops.

\subsubsection*{Step 5b. Emission into veto region}
If the emission $n_k$ lands in the gap region, the shower stops.

\section{Fixed-order expansion of the LL result \label{sec:FOofLL}}
In this appendix, we detail how one can obtain the coefficients $\mathcal{S}^{(1)}$ and $\mathcal{S}^{(2)}$ in the fixed-order expansion of the leading logarithmic resummation \eqref{eq:fixedorderexpansion} from the parton shower. When extracting the fixed-order coefficients for a given tree-level event, we look at each dipole at $t=0$ individually, calculate its expansion coefficients and then combine the results. Averaging the expansion coefficients of the individual events then gives the final result for the two-loop expansion of the resummed cross section. In the following we describe the computation for a single dipole.

\subsection{One-loop coefficient}
The one-loop coefficient of the fixed-order expansion may be extracted easily from the shower. To do so we write \eqref{eq:Sone}  as 
\begin{align}\label{eq:app_s1rewrite}
\mathcal{S}^{(1)} &= - V_{12} \,\int \frac{d\Omega (n_3)}{4\pi}  \frac{R_{12}^{3}}{V_{12}} \Theta_{\rm out}(n_3)\,,
\end{align}
where $1$ and $2$ are the legs of the dipole and $R_{12}^{3} = 4N_c \widetilde{W}_{12}^3$. Please note that throughout this appendix we write out the appropriate $\Theta_{\rm in}(n_k)$ and $\Theta_{\rm out}(n_k)$ angular constraints and we do not include the factors $\Theta_{\rm in}(n_k)$ into the definition of $R_{ij}^{k}$ as we did in \eqref{eq:app_realEmission}.

The factor ${R_{12}^{3}}/{V_{12}}$ is produced by the shower, see Appendix \ref{sec:MCalg}, and gives $\mathcal{S}^{(1)}$ after multiplication by $-V_{12}$. All that needs to be done is to account for the constraint $\Theta_{\rm out}(n_3)$ which ensures that the emission is in the veto region. From the first step of the parton shower one obtains
\begin{align}
\mathcal{S}^{(1)} &= -\sum_{i=1}^{n_{\rm{sh}}} s_1\,,\hspace{1cm}\text{with}\hspace{1cm} s_1\equiv \begin{cases}
\frac{R_{12}^{3}}{V_{12}}\frac{V_{12}}{n_{\rm sh}},& \text{if } \Theta_{\rm out}(n_3)=1\,,\\
0,              & \text{otherwise.}
\end{cases}
\end{align}

\subsection{Two-loop coefficient}
While the global part of the two-loop coefficient is just one half of the one-loop coefficient squared, the non-global part is much more involved mainly due to collinear divergences in individual terms in the integrand in \eqref{eq:fixedordercoeff}. Rewriting the non-global part as
\begin{align}
\mathcal{S}_{\rm NGL} ^{(2)}
&=-\frac{1}{2}  \int \frac{d\Omega (n_3)}{4\pi}\int \frac{d\Omega (n_4)}{4\pi}\,   R_{12}^{3} \left(\, R_{13}^{4}+ R_{23}^{4} - R_{12}^4 \right)\Theta_{\rm in}(n_3)\Theta_{\rm out}(n_4)\, ,
\end{align} 
one notices that collinear singularities arise in $R_{12}^{3}$ since the light-like direction $n_3$ is in the jet region (not in the veto region) and can become collinear to $n_1$ or $n_2$. However, the full expression $\mathcal{S}_{\rm NGL} ^{(2)}$ is collinear finite since the terms multiplying $R_{12}^{3}$ vanish in both collinear limits:
\begin{equation}
R_{13}^{4}+ R_{23}^{4}- R_{12}^4 \to 0 \quad \text{ for } \quad n_3 \to n_1  \quad \text{ or } \quad n_3 \to n_2\,.
\end{equation}

The two terms $R_{12}^{3} R_{13}^{4} $ and $R_{12}^{3} R_{23}^{4}$ have a simple interpretation in the parton shower. The first emission of the shower produces $R_{12}^{3}$ and results in the dipole configuration $(1,3,2)$. The term $R_{13}^{4} $ arises when the second emission occurs in the dipole $(1,3)$, while $R_{23}^{4}$ corresponds to the emission from $(3,2)$. However, the subtraction term $R_{12}^4$ does not arise in the parton shower. To include it, the factor $R_{12}^4$  can be split in two parts according to
\begin{align}
R_{12}^{4}=R_{12}^{4}\,\theta(n_2\cdot n_3-n_1\cdot n_3)+R_{12}^{4}\,\theta(n_1\cdot n_3-n_2\cdot n_3) \, .
\end{align}
In this way, the first term is evaluated if the spatial angle between $n_3$ and direction $n_1$ is smaller than the one between $n_3$ and direction $n_2$ and is therefore used to cure the collinear singularity when $n_3\rightarrow n_1$ .The same argument holds for $1\leftrightarrow 2$ and removes the other divergence. One can thus write 
\begin{align}\label{eq:S2split}
\mathcal{S}_{\rm NGL} ^{(2)}&=-\frac{1}{2}  \int \frac{d\Omega (n_3)}{4\pi}\int \frac{d\Omega (n_4)}{4\pi}\, \Theta_{\rm in}(n_3)\Theta_{\rm out}(n_4)\nonumber\\
&\hspace{3cm} \times \Bigg[\, R_{12}^{3}R_{13}^{4}\left(1-\frac{R_{12}^4}{R_{13}^4}\theta(n_1\cdot n_3-n_2\cdot n_3)\right)\nonumber\\
&\hspace{4.5cm}+ R_{12}^{3}R_{23}^4\left(1-\frac{R_{12}^4}{R_{23}^4}\theta(n_2\cdot n_3-n_1\cdot n_3)\right)\Bigg]\, .
\end{align} 
The terms in the second and third line of this expression are separately collinear finite and the factors multiplying the dipoles $R_{12}^{3}R_{13}^{4}$ and $R_{12}^{3}R_{23}^4$ in the two lines can be implemented as weight factors in the shower.

To implement \eqref{eq:S2split} in the  shower, we store the weight  $R_{12}^{3}/V_{12}$ of the first emission. We then go on to the second emission and check whether it is emitted by the dipole $(n_1,n_3)$ or $(n_3,n_2)$. We also check if $n_3$, the direction of the first emission, is closer to $n_1$ or $n_2$. With this information, one can then  calculate the two-loop coefficient as a sum of weights 
\begin{equation}
\mathcal{S}_{\rm NGL} ^{(2)}=-\frac{1}{2}\sum_{i=1}^{n_{\rm{sh}}} s_2\,.
\end{equation}
The weights for the two cases $(i,j)=(1,2)$ and $(i,j)=(2,1)$, corresponding to the second emission $R_{i3}^{4}$, are obtained as follows
\begin{align}
s_2&\equiv \begin{cases}
\frac{R_{12}^{3}}{V_{12}}\frac{V_{12}}{V_3}\frac{V_{i3}}{V_3}\frac{R_{i3}^{4}}{V_{i3}}\frac{(V_3)^2}{n_{\rm{sh}}}\left(1-\frac{R_{12}^{4}}{R_{i3}^{4}}\right) ,& \text{if $\Theta_{\rm out}(n_4)=1$ and $n_i\cdot n_3>n_j\cdot n_3$\, ,}\\
\frac{R_{12}^{3}}{V_{12}}\frac{V_{12}}{V_3}\frac{V_{i3}}{V_3}\frac{R_{i3}^{4}}{V_{i3}}\frac{(V_3)^2}{n_{\rm{sh}}}\, ,& \text{if $\Theta_{\rm out}(n_4)=1$ and $n_i\cdot n_3<n_j\cdot n_3$\, ,}\\
0\,,              & \text{otherwise}\, .
\end{cases}
\end{align}
We have written $s_2$ in terms of factors ${V_{12}}/V_3$ and ${V_{i3}}/V_3$, with $V_3=V_{13}+V_{32}$, which arise in the shower algorithm, analogous to \eqref{eq:H5HatT}. They represent the probability to continue the shower after the first emission and the probability to choose the dipole $(n_i,n_3)$ rather than $(n_j,n_3)$ for the second emission.

\end{document}